\documentclass[useAMS,usenatbib]{mn2e}
\usepackage{epsfig,appendix,array,rotating,pdflscape,array,longtable,colortbl}
\usepackage{multirow,url}
\usepackage{booktabs,graphicx}
\usepackage{lscape, amsmath} 
\usepackage{paralist}
\usepackage{relsize}
\usepackage{fixltx2e}
\usepackage{amssymb}
\newcommand{\hi}{H\,{\sc i}}
\newcommand{\hii}{H\,{\sc ii}}

\newcommand{\km}{km\,s$^{-1}$}
\newcommand{\degree}{$^{\circ}$}
\newcommand{\halpha}{H${\alpha}$}
\newcommand{\cg}{CGCG\,97-}
\newcommand{\msolar}{M$_{\odot}$}

\newcommand{\hdos}{H$_{2}$}

\newcommand{\angs}{${\mathrm {\AA}}$}

\newcommand{\hto}{CO ($J=2 \rightarrow 1$)}
\newcommand{\hoz}{CO ($J=1 \rightarrow 0$)}
\newcommand {\apgt} {\ {\raise-.5ex\hbox{$\buildrel>\over\sim$}}\ }
\newcommand {\aplt} {\ {\raise-.5ex\hbox{$\buildrel<\over\sim$}}\ }

\setlongtables
\title[]{ \textcolor{black}{Highly perturbed molecular gas in infalling cluster galaxies: the case of CGCG97-079} }
\author[]{T. C. Scott$^{1,2}$\thanks{E-mail: tom.scott@astro.up.pt (TCS)}, A. Usero$^{3}$, E. Brinks$^{2}$, H. Bravo--Alfaro$^{4}$, L. Cortese$^{5}$, and  A. Boselli$^{6}$,\and 
and M. Argudo--Fern\'andez$^{7,8,9}$
\\
$^{1}$\textcolor{black}{Institute of Astrophysics and Space Sciences (IA), Rua das Estrelas, 4150-762 Porto, Portugal}\\
$^{2}$Centre for Astrophysics Research, University of Hertfordshire, College Lane, Hatfield, AL10 9AB, UK \\
$^{3}$Observatorio Astron\'omico Nacional, C/Alfonso XII 3, 28014 Madrid, Spain\\
$^{4}$Departamento de Astronom\'\i a, Universidad de Guanajuato, Apdo.\ Postal 144, Guanajuato 36000, Mexico\\
$^{5}$Centre for Astrophysics and Supercomputing, Swinburne University of Technology, PO Box 218
Hawthorn, Victoria 3122, Australia\\
$^{6}$Aix Marseille Universit\'e, CNRS, LAM (Laboratoire d'Astrophysique de Marseille) 
UMR 7326, 13388, Marseille, France\\
$^{7}$Instituto de Astrof\1sica de Andaluc\1a (CSIC), Apartado 3004, 18080 Granada, Spain\\
$^{8}$Departamento de F\1sica Te\'orica y del Cosmos, Universidad de Granada, 18071
Granada, Spain\\
$^{9}$Key Laboratory for Research in Galaxies and Cosmology, Shanghai Astronomical Observatory, Chinese Academy of Sciences, \\
80 Nandan Road, Shanghai, China, 200030}

\begin{document}
\date{Accepted }

%\pagerange{\pageref{firstpage}--\pageref{lastpage}} \pubyear{2008}

\maketitle

\label{firstpage}

\begin{abstract}
\textcolor{black}{We report on  \hto\  mapping with the IRAM 30--m HERA receiver array of \cg079, an irregular galaxy in the  merging galaxy cluster Abell\,1367 (\textit{z} = 0.022).   We find that $\sim$ 80\% of the detected \hto\ is projected within a 16 arcsec$^{2}$ (6.5 kpc$^{2}$) region to the north and west of the optical/NIR centre, with the  intensity maximum offset $\sim 10$\,arcsec (4\,kpc)   NW of  the  optical/NIR centre and  $\sim$  7 arcsec (3 kpc) south--east of the \hi\ intensity maximum. Evolutionary synthesis models  indicate  \cg079 experienced a burst of star formation $\sim$ 10$^8$ yr ago, most likely triggered by a tidal interaction with \cg073. For \cg079 we deduce  an infall velocity to the cluster of $\sim$ 1000 \km\ and  moderate ram pressure  (P$_\mathrm{ram} \approx 10^{-11}$\,dyn\,cm$^{-2}$).   The observed offset in \cg079  of the highest density \hi\ and  \hto\ from  the stellar components  has not previously been observed in galaxies \textcolor{black}{currently undergoing} ram pressure stripping, although previous detailed studies of  gas morphology and kinematics during ram pressure stripping were restricted to  significantly more massive galaxies with deeper gravitational potential wells.  We conclude the observed  \textcolor{black}{ cold gas} density maxima offsets are most likely  the result of ram pressure and/or  the  high--speed tidal  interaction with \cg073.  However ram pressure stripping is likely to be  playing a major role in the perturbation of lower density  gas.}

\end{abstract}


\section{INTRODUCTION}
\label{into}
\textcolor{black}{The consensus view is that, in low-redshift galaxy clusters, ram pressure stripping by the intra--cluster medium (ICM) is the
dominant mechanism accelerating the evolution of late--type galaxies \citep{vgork04,bosel14b,chung09}. Models for cluster spirals subject to ram pressure stripping \citep[e.g.,][]{roed05,roed07,kapf09,tonn09}  indicate  that removal of the interstellar medium (ISM) proceeds progressively during a spiral's infall to the cluster. Initially only the loosely bound peripheral interstellar medium (ISM), principally \hi, is stripped. The higher
column density \hi\  and molecular gas located in the deepest
parts of a galaxy's potential well, \textcolor{black}{traced by  the optical/Near 
Infared (NIR) intensity maxima,} remain unperturbed until late in
the stripping process and requires  high velocities relative to the ISM/ICM 
(V$_\mathrm{rel}$) and high ICM densities encountered during a galaxy's  traverse of a cluster 
core in order to remove them \cite[e.g.,][]{roed05,roed07}.  \textcolor{black}{The truncated \halpha\ and dust  disks observed in  \hi\ deficient spirals are consistent with this picture of outside--in ISM  removal \citep{koop04,cort10}.   } \hi\ and CO observations
and modelling of the Virgo spirals NGC\,4522 and NGC\,4330
by \cite{voll08,voll12} are examples showing that during ram pressure stripping the high column density \hi\
and molecular gas remain located  deep within the gravitational
well traced by the NIR and optically most luminous part of the
galaxies.}  \textcolor{black}{Moreover NGC 4848 \citep{voll01b} in Coma \citep[$L_X$ =7.8 x 10$^{44}$ ergs s$^{-1}$;][]{plionis09}  indicates that even in X--ray luminous clusters a spiral's transit of the cluster core may remove  \hi\  without causing a molecular gas deficiency\footnote{\textcolor{black}{A spiral's \hdos\ or \hi\ deficiency is defined as the log of the ratio of the expected to observed gas mass. Negative values indicate an excess. }}, although the recent work by \cite{bosel14} for a sample of Virgo galaxies shows a correlation between molecular and \hi\ deficiencies \textcolor{black}{ \cite[see also][]{fuma09}}.  The ISM content, morphology and kinematic signatures resulting from extreme ram pressure ($\sim$ 10$^{-10}$ dyne cm$^{-2}$),  as is proposed for  ESO\,137--001 projected $\sim$ 280 kpc from the center of the Norma cluster ($M_{dyn}$ $\sim 1 \times 10^{15}$\,\msolar)   \textcolor{black}{from modelling with } $V_\mathrm{rel}$ $\sim$ 3000 \km\ by  \cite{jachym14}, are less well studied.   Those  authors argue extreme  ram pressure stripping has already removed all detectable \hi\ from the disk of ESO\,137--001 and is now in the process of stripping the molecular gas, as well as being partially responsible for a molecular tail containing approximately half of the galaxy's observed molecular gas. \cite{jachym14} note  the stripped  \hi\ is likely to have have been converted to other phases, \textcolor{black}{including the observed  molecular and X--ray tail emission}. Even ESO\,137--001 appears to have followed the sequential removal of increasingly dense ISM  from the galactic disk  during ram pressure stripping\textcolor{black}{, with} stripping of the molecular disk  being delayed until the final stages which are now being witnessed.} 

\textcolor{black}{There have been relatively few detailed studies of changes in \hi\ and CO morphology and gas kinematics during ram pressure stripping in gas--rich late--type spirals and \textcolor{black}{in particular of} dwarf irregular galaxies.  \textcolor{black}{Although a Virgo Irr dwarf, IC\,3418, with about twice the stellar mass of the SMC, contains evidence of having} lost almost all its \hi\ and molecular gas by ram pressure stripping within the last few $\times$ 10$^8$ yr  \citep{ken14}.  } A sample of late--type dwarf galaxies near the centres of clusters, which are  presumed to have lost their \hi\  by ram pressure stripping,  display truncated \halpha\ disks   \citep{fosssati13}. \textcolor{black}{Also we see clear evidence of truncated  UV star-forming disks in  both \textcolor{black}{spiral} and dwarf galaxies in the Virgo cluster \citep{cort12}.} \textcolor{black}{These studies imply} dwarfs are subject to  the same sequential  outside--in  gas stripping as higher mass late--type galaxies \citep{bosel14b}, \textcolor{black}{a proposition  which is supported by the detection of CO in the disk of IC\,3418.}   

\textcolor{black}{The scenario depicted above \textcolor{black}{is  challenged by our recent
studies of late--type galaxies in the spiral-rich merging cluster 
Abell\,1367 \citep[$L_{X}$ = 1.25 $\times$ 10$^{44}$  egs s$^{-1}$;][]{plionis09} }  which has a $M_{dyn}$  $\sim 6.9 \times 10^{14}$\,\msolar\  \citep{bosel06a}, about half that of Coma \textcolor{black}{or Norma}, and hosts two sub--clusters in the early stages of an approximately equal mass merger  \citep{don98}. 
In \citet[][hereafter Paper I]{scott10}, we found an unusually high frequency of offsets between 
the \hi\ and optical intensity maxima in a sample \textcolor{black}{of the cluster's}  late--type galaxies. 
\textcolor{black}{Our multi--pointing 
\hoz\ and \hto\ observations with the IRAM\footnote{Institut de radioastronomie millim\'etrique (IRAM) is supported by CNRS/INSU (France), the MPG (Germany) and the IGN (Spain).} 30--m telescope \citep[][hereafter Paper II]{scott13},}  provided \textcolor{black}{indications} of perturbed molecular disks and molecular gas excesses in \textcolor{black}{a} subset of the Paper I sample. 
Together, these papers  raise the question whether a mechanism other than standard ram pressure stripping \textcolor{black}{is driving} the evolution of spiral galaxies in A\,1367, and in merging clusters in general.}

\begin{figure*}
\begin{center}
\includegraphics[ angle=0,scale=0.5] {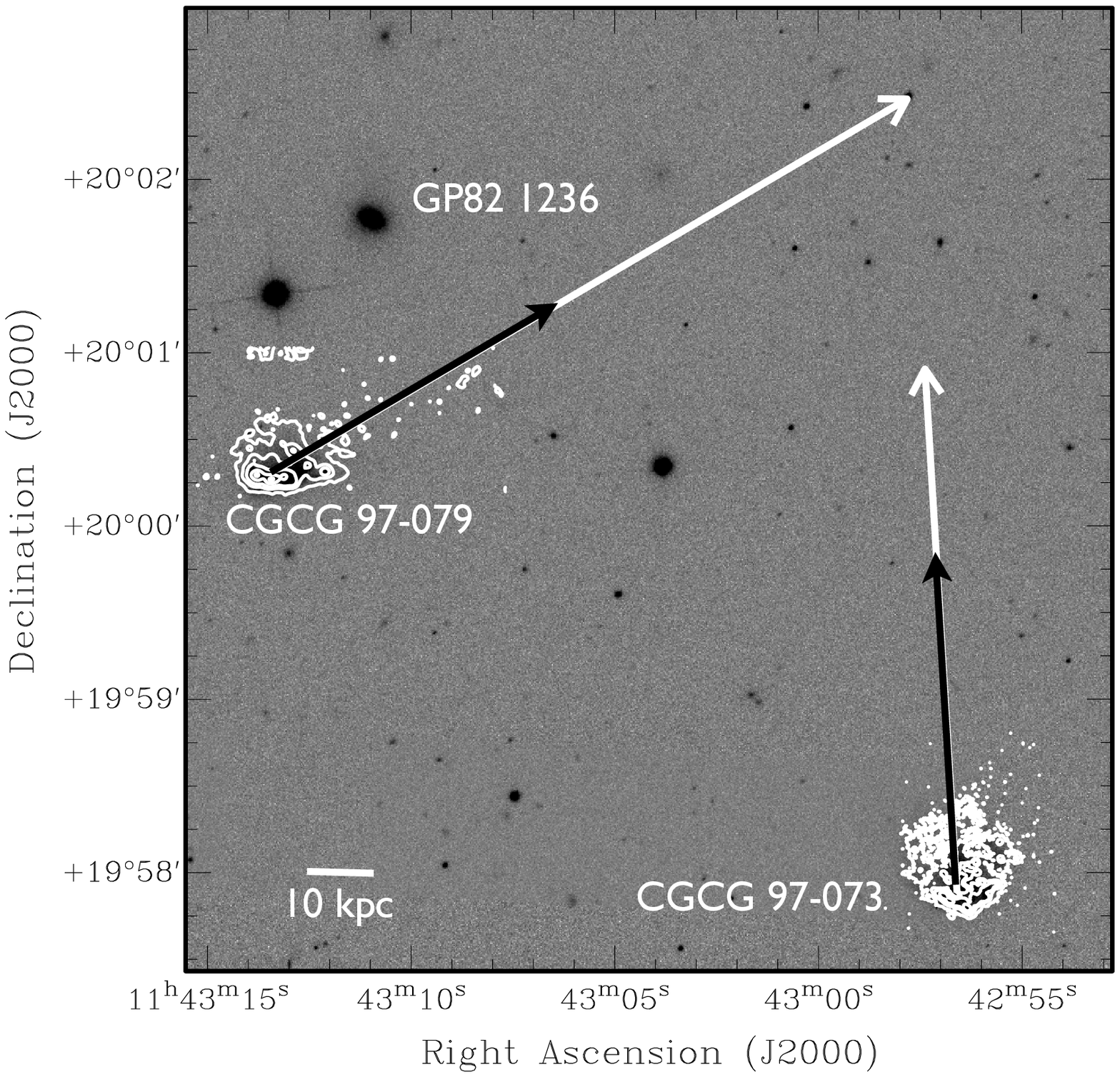}
\vspace{1cm}
\caption{\cg079: SDSS \textit{r}--band image with the \textcolor{black}{black (shorter) and white (longer)}  arrows indicating, respectively, the approximate extent and orientation of the  radio continuum \citep{gava87} and \halpha\ tails \textcolor{black}{of \cg073 and \cg079}. The  white arrow \textcolor{black}{ head of the \cg079 tail marks the approximate position where the \halpha\ tails from deep imaging from Boselli \& Gavazzi (2014) appear to } meet.  \ \halpha\  contours (\textcolor{black}{white}) are from a GOLDMine image. The position of the nearest neighbouring galaxy,  A\,1367 GP82 1236 \textcolor{black}{ is also marked}.}
\label{overv}
\end{center}
\end{figure*}

\textcolor{black}{ \cg079 is an optically--irregular galaxy with $D_{25}$ = 45 arcsec (18.6 kpc), an estimated $M_{*}$  $\sim 1.2 \times 10^{9}$\,\msolar, (i.e., of order of the Large Magellanic Cloud or M33) and is projected 420\,kpc to the NW of the NW sub--cluster core  of A\,1367 (Figure 1 in Paper II). \textcolor{black}{That figure also shows the X--ray emission (\textit{ROSAT}) from the A\,1367 ICM.  \textcolor{black}{Its} 50--\textcolor{black}{100} \,kpc ionised gas and radio continuum tails oriented away from the cluster centre (Figure  \ref{overv}), provide evidence that \cg079 is undergoing} ram pressure stripping \citep{gava87,gava01b,bosel14b}. However \textcolor{black}{our previous observations (Papers I and II) suggest}  its neutral (atomic and molecular) gas distributions are inconsistent with current ram pressure stripping models. In particular its molecular and atomic gas intensity maxima lie $\sim$6 kpc and $\sim$12\,kpc, respectively, to the NW from its optical intensity maximum as derived from sparsely sampled CO maps and low resolution \hi\ data. \textcolor{black}{ \cg079 is the clearest case in A\,1367 of \hi\ and CO intensity maxima  offsets, in apparent contradiction to ram pressure models,  making it  \textcolor{black}{an  interesting}  target for the investigation of mechanisms driving the  the evolution of  ISM in galaxies in A\,1367.}
}

 This \textcolor{black}{paper} reports on follow--up  \hto\ mapping of \cg079  with the IRAM 30--m HERA receiver array.  \textcolor{black}{We also utilised  \hi\ C--array data from the NRAO\footnote{The National Radio Astronomy Observatory is a facility of the National Science Foundation operated under cooperative agreement by Associated Universities, Inc.} VLA archive (project AG264), a \textit{Spitzer} \textcolor{black}{IRAC}\footnote{InfraRed Array Camera.} 4.5\,$\mu$m image (ID: AORKEY25789696 taken from the Infrared Science Archive IRSA), a {\em GALEX} FUV image ({A1367\_SPEC\_A}-fd-int.fits) retrieved from the MAST archive \footnote{\textcolor{black}{Mikulski Archive for Space Telescopes.}}, and  an \halpha\  image from GOLDMine\footnote{ Galaxy Online Database Milano Network,\citep[][http://goldmine.mib.infn.it/]{gava03b}. } }

 Section \ref{allobs} gives details of the \textcolor{black}{HERA} observations,  with  observational results in section \ref{results}.  A discussion follows in section \ref{discuss} with concluding remarks in section \ref{concl}. Based on a redshift to A\,1367  of 0.022 and assuming $\Omega_{\mathrm m} =0.3$, $\Omega_\Lambda =0.7$, and  $H_o  =72$\,\km\,Mpc$^{-1}$ \citep{sperg07} the distance to the cluster is 92\,Mpc and the   angular scale is 1\,arcmin $\sim$ 24.8\,kpc. All equatorial positions referred to throughout this paper are in J2000.0. \textcolor{black}{References to the optical centre are to the position 11$^{\mathrm h}$43$^{\mathrm m}$13\fs4, +20$\degr00^\prime17^{\prime\prime}$  from the NASA/IPAC Extragalactic Database (NED)}.

\section{OBSERVATIONS--HERA CO MAPPING }
\label{allobs}

$^{12}$\hto\  emission line mapping of \cg079 was carried out with the HERA $3 \times 3$ multi--beam receiver array on the  IRAM  30--m telescope at Pico Veleta, Spain, in \textcolor{black}{slightly  under sampled} dual polarisation ``On the Fly'' mode with  position switching. \textcolor{black}{Receivers used were HERA 1 and  2 with  WILMA 1 and  2 backends (1024 channels with  2 MHz separation).}  Further observational parameters are given in Table~\ref{obs}. The total integration time was 13\,hr which was accumulated in March 2010 and January 2012.  \hto\ was mapped  in a $\sim1.1\times 1.1$\,arcmin$^2$  region of sky centred at  (-10.6, +10.6 arcsec) from the galaxy's optical centre.   After every three scans ($\sim 15$ minutes) a chopper wheel calibration was carried out. Pointing was checked at between 1 and 2 hour intervals using the broadband continuum of \textcolor{black}{3C273}.  

\begin{table} 
\centering
\begin{minipage}{140mm}
\caption{HERA observational parameters}
\label{obs}
\begin{tabular}[h]{@{}l@{}l@{}l@{}r@{}}
\hline
Rest frequency& $^{12}$\hto\ & $[$GHz] &230.5 \\
Sky frequency &&$[$GHz] &225.3\\
Primary Beam&FWHP\footnote{Full width half power}&[arcsec]&11.5\\
Primary Beam&FWHP&[kpc] &4.5\\
Receivers &HERA && 1 \& 2 \\
Backend channel width&&[MHz]& 2\\
T$_\mathrm{sys}$ &&[K]&300$\pm$100\\
$\tau$ @ 225 GHz &&&0.3$\pm$0.1\\
\textcolor{black}{Expected pointing accuracy}&&\textcolor{black}{[arcsec]}&\textcolor{black}{2}\\
T$_\mathrm{mb}$&&&1.73 $T_{A}$*\\
\hline
\end{tabular}
\end{minipage}
\end{table}

Data reduction was carried out using GILDAS software and  excluded spectra taken under particularly poor conditions. Spectra were then summed position by position.    The data cube was produced using the {\sc PLAIT} task and excluded  \textcolor{black}{unevenly--sampled }edge regions. Observations covered the velocity range 6552 \km\ to 7425 \km\ with a velocity resolution of 11 \km. The  data cube was blanked using the AIPS\footnote{Astronomical Image Processing System} software package task {\sc blank} resulting in a final cube containing 42  channels binned to a velocity resolution of 21\,\km\ \textcolor{black}{with an}  angular resolution of 11.5 \textcolor{black}{arcsec.}  

\section{OBSERVATIONAL RESULTS}
\label{results}

\begin{figure*}
\begin{center}
\includegraphics[ angle=0,scale=.55] {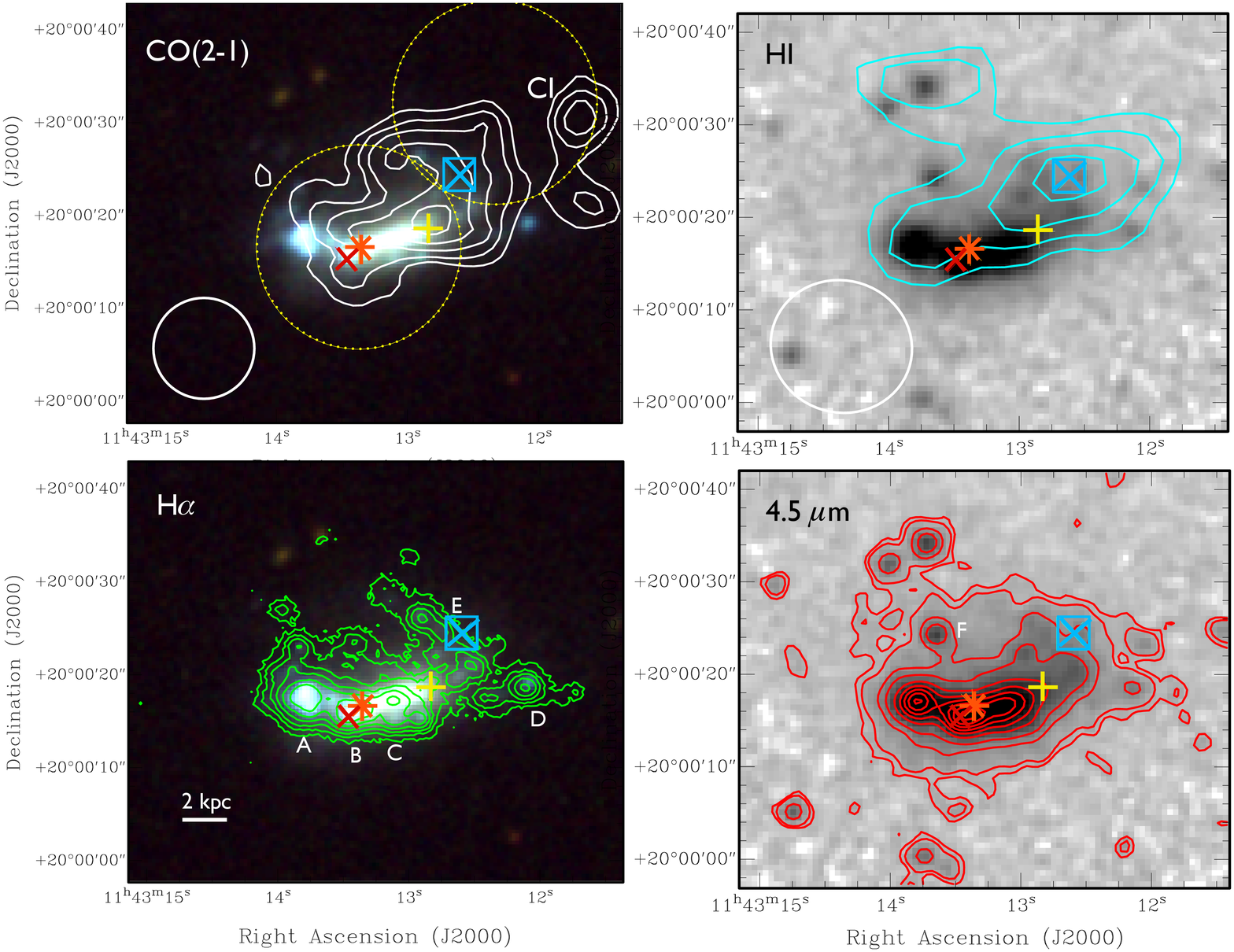}
\vspace{1cm}
\caption{\cg079 \textit{top left:}  \hto\ HERA \textcolor{black}{integrated intensity map} contours \textcolor{black}{(gray)} overlaid on  a composite  SDSS  false colour $g, r, i$ -- band image.   The first contour level is at 3 $\sigma$, equivalent to \textcolor{black}{0.68 K \km\ (T$_\mathrm{mb}$) and subsequent contours are at 1.0, 1.3, 1.7, and 2.0  K \km\ (T$_\mathrm{mb}$).  The white circle indicates the size of the \hto\ beam and CI indicates the position of the detached clump. \textcolor{black}{The IRAM 30--m 22 arcsec FWHP single beams for the \hoz\ Paper I  observations at the optical centre and offset $RA$ = -15.6 arcsec, $DEC$ =15.6 arcsec are indicated with large dashed yellow circles.  }}  \textit{Top right:} \textcolor{black}{ \hi\ (VLA C--array) contours (cyan) overlaid on  a \textcolor{black}{4.5 $\mu$m image  \textcolor{black}{(resolution $\sim$ 2 arcsec)}}. The contours are at 2, 2.5, 3, and 3.5 $\sigma$.  The white circle indicates the size of the VLA C--array beam.   \textit{Bottom  left:} }   \halpha\ contours (green) from GOLDMine   \textcolor{black}{(resolution $\sim$ 3 arcsec)}  overlaid on   a composite SDSS  \textcolor{black}{ false colour $g, r, i$ --} band image. \textcolor{black}{The contours are at 20, 50, 100, 200, 400 and 800 $\sigma$.   } The letters A to E mark the 5 principal \hii\ regions (\halpha\ knots), following the naming convention from Figure \ref{fig2} of \citep{gava95}.  \textit{Bottom right:}  \textit{Spitzer} \textcolor{black}{contours from a 4.5  $\mu$m boxcar smoothed image, overlaid on a 4.5  $\mu$m un--smoothed image. The contours are at 4.5, 5.5, 10, 15, 25, 35, 45, 55, 65  and 75  $\sigma$ with the 4.5 $\sigma$ contour equivalent to  0.085 MJy sr$^{-1}$. The resolution of the 4.5 $\mu$m image is $\sim$ 2.5 arcsec.}  The \textcolor{black}{orange asterisk marks the optical centre  \textcolor{black}{and  the cyan 'X' in box marking the position of the \hi\ intensity maximum from the VLA C--array}. The red 'X' corresponds to the 4.5  $\mu$m} maximum and the \textcolor{black}{yellow} plus symbol the \hto\  maximum from the HERA map. } 
\label{fig2}
\end{center}
\end{figure*}

\begin{figure*}
\begin{center}
\includegraphics[ angle=0,scale=0.50] {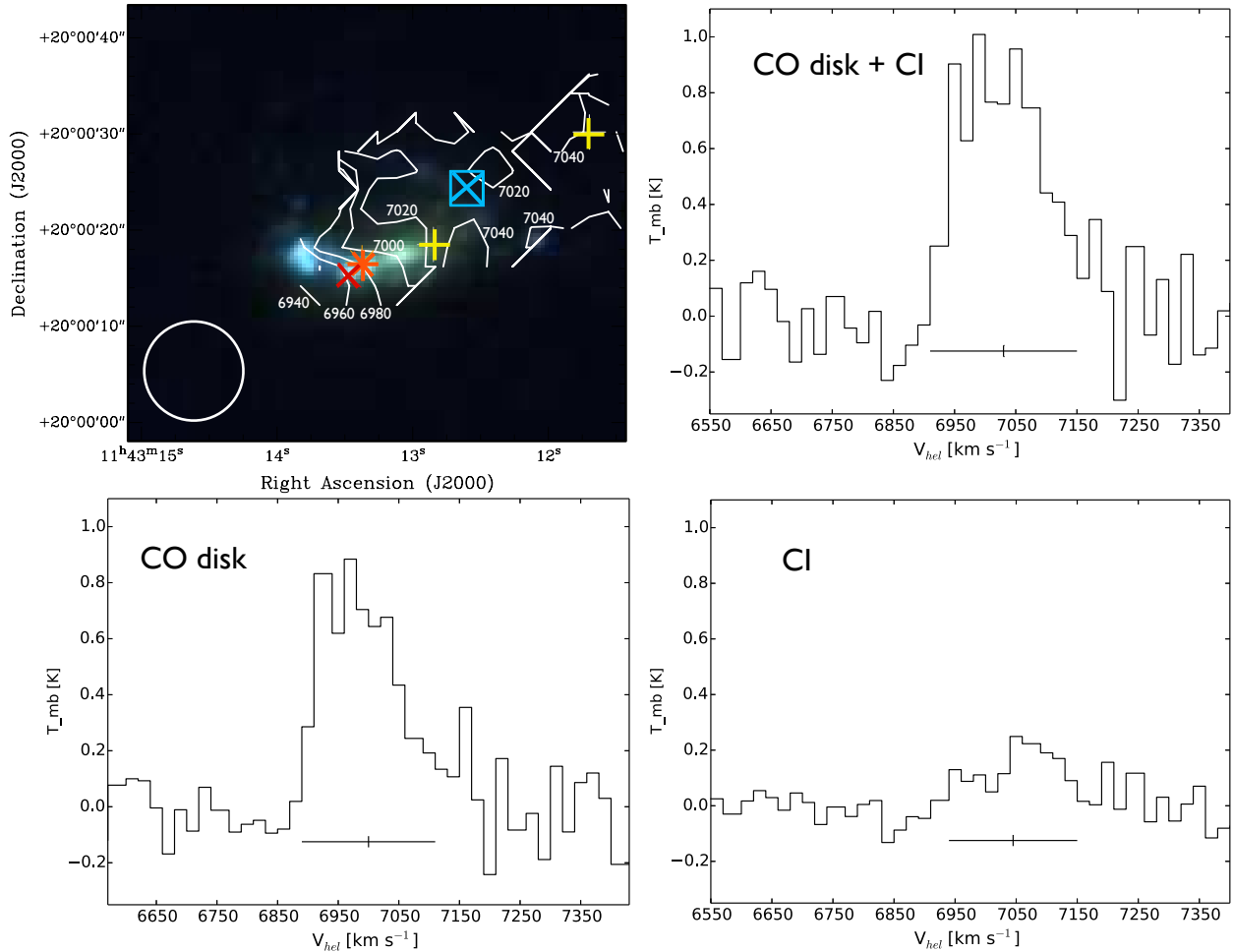}
\vspace{1cm}
\caption{ \textcolor{black}{\textit{Top left panel:} \hto\ velocity map with contours (\km) overlaid on a composite SDSS -- $g, r, i$ band image. The white circle at the bottom left of the panel  shows the approximate size of the HERA \hto\ beam. The colours and symbols  are as in Figure \ref{fig2} except the \textcolor{black}{yellow plus symbol at} the \textcolor{black}{top right which marks CI.} \textit{Top right panel:}  HERA \hto\ spectra for the CO disk + CI. \textit{Bottom left panel}:  HERA \hto\ spectrum for the CO disk.   \textit{Bottom right panel}: HERA \hto\  spectrum for CI.  \textcolor{black}{ The T$_\mathrm{mb}$ for each channel in  a spectrum (21 \km\ velocity width per channel) is the sum of the  $T_{A}$* for all pixels in the channel, converted to the T$_\mathrm{mb}$ scale.}  The horizontal line at the bottom of each spectrum indicates  the \textcolor{black}{spectrum's W$_{20}$ velocity width } and the vertical cross bar the V$_{hel}$.  } }
\label{fig_3}
\end{center}
\end{figure*}

 \textcolor{black}{The \hto\ contours from the HERA \textcolor{black}{integrated intensity map}, in  Figure~\ref{fig2} (top left),  show that the bulk of  \hto\ was  detected within a continuous emission region approximately 16\,arcsec$^2$  (7\,kpc$^2$)} in area,   projected  west and  north of the optical centre.  This continuous emission region is  referred to as the ``CO disk".   The CO disk contains the \hto\  intensity maximum (CO maximum), indicated with  \textcolor{black}{yellow} plus signs in Figure~\ref{fig2},  and is  located $\sim 9.9$\,arcsec (4.1\,kpc) NW of the \textit{Spitzer } 4.5  $\mu$m intensity maximum ($PA$ = 287\degree). \textcolor{black}{Also marked in  the same figure is a detached \hto\ clump (CI)  $\sim$ 29 arcsec (12 kpc)  NW of the optical centre.  CI is detected in 3 contiguous channels at $>$  3 $\sigma$. We are  cautious  about declaring CI a detection because of its low peak signal to noise (S/N) of \textcolor{black}{3.9}\textcolor{black}{, its lack of a counterpart at any other wavelength}   and its proximity to the map edge. However on balance its \hto\  spectrum (Figure \ref{fig_3} -- bottom left), \textcolor{black}{\hto\ single beam detection from Paper II, }velocity and the total \hdos\  mass  including CI (see next paragraph) support CI  being a real feature.} The CO disk morphology of \cg079 differs dramatically from that  of local spirals \citep{nishiyama01,leroy09}. If it  followed that of the HERACLES nearby spiral sample \citep{leroy09}, it would  display an exponential distribution with its origin at the optical centre with a   \textcolor{black} typical expected scale length of 0.23\,$r_{25}$ (i.e., 3.4\,arcsec or 1.4\,kpc).  

L(CO)$_{2-1}$   \textcolor{black}{from the CO disk and CI is \textcolor{black}{1.4}$\pm0.1\times 10^8$ K \km pc$^2$ (T$_\mathrm{mb}$  scale) which  converted  to an \hdos\ mass, using equation} B11\footnote{ Equation B11 requires L(CO)$_{1-0}$  and to apply the equation we used the conversion:  \textcolor{black}{$R21/R10$ = 0.75}} from Paper II ($X$ factor of $2.62 \times 10^{20}$\,molecules\,cm$^{-2}/$K\,km\,s$^{-1}$), gives $M(H_2$) = \textcolor{black}{0.8} $\pm0.1 \times10^9$ \msolar. \textcolor{black}{ This compares to the earlier  \hdos\ mass  estimate based on single beam \hoz\ pointings  (Paper~II)   of 1.07 $\times10^9$ \msolar. } \textcolor{black}{ Figure \ref{fig2} -- top left shows the position of the \hoz\  full width half power (FWHP) single beam pointings (dashed yellow circles), from which the earlier \hdos\ mass was estimated. The HERA  data presented here have a higher signal to noise.  A comparison of spectra extracted from the current data with the single \hto\  pointings from Paper ~II leads us to suspect the Paper II  masses are less reliable.} Further  properties of  the CO disk and CI  are given in Table~\ref{hera} with V$_\mathrm{hel}$, and W$_{20}$ calculated using the methods from Paper~II.
%The  HERA  integrated velocity spectrum for the  CO disk   has a single peak rather than the double horned profile of an edge--on  rotating  gas disk  expected to coincide with either an inclined optical disk proposed by \cite{gava95} or  the infared disk defined in section \ref{results_NIR}.  
 
\begin{figure*}
\begin{center}
\includegraphics[ angle=0,scale=0.8] {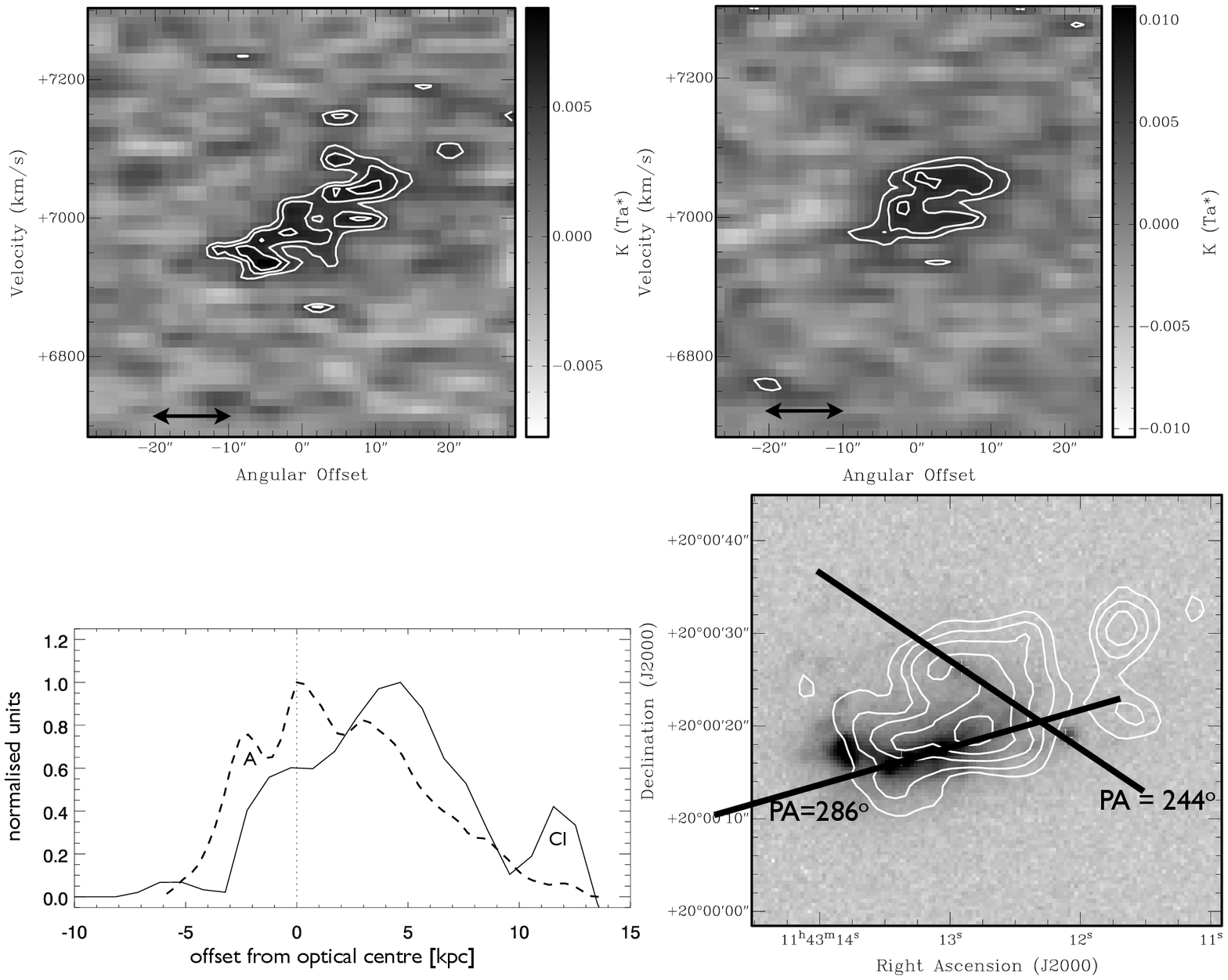}
\vspace{1cm}
\caption{ \textcolor{black}{\textit{Top left panel:} PV diagram ($PA$ = 286\degree) centred at the optical centre.  The angular offset, in arcsec, is negative to the SE (left)  and positive to the NW (right). \textit{Top right panel:}  PV diagram \textcolor{black}{($PA$ =244\degree) along the line connecting knot E and D centred at knot E}. The angular offset, in arcsec, is negative to the NE (left)  and positive to the SW (right). The horizontal arrows indicated the size of the HERA beam. \textit{Bottom \textcolor{black}{left} panel}: }\cg079  integrated \textit{Spitzer} IRAC 4.5 $\mu$m (\textcolor{black}{dashed line}) and \hto\ (\textcolor{black}{solid line}) emission \textcolor{black}{in normalised units} along the  $PA$ = \textcolor{black}{286}\degree\ centred on the optical centre of the galaxy. The horizontal axis shows the offset from the optical centre in kpc, with positive values to the NW.  \textcolor{black}{\textit{Bottom right panel: }SDSS g -- band image showing the orientation and lengths of the slices along which \textcolor{black}{the} PV diagrams where derived.} }
\label{fig_4}
\end{center}
\end{figure*}

 \textcolor{black}{Figure \ref{fig_3} (top left) shows the \hto\ velocity field for the CO disk and CI in the regions where emission is $>$ 3 $\sigma$ in the integrated intensity  map. From this velocity field and PV diagram (Figure \ref{fig_4} -- top left) we see a systematic increase in velocity from 6960 \km\ along an axis joining the optical centre and the  \hto\ intensity maximum  where the  velocity reaches 7020 \km. From the \hto\  maximum the velocity remains more or less constant between 7020 \km\  to 7040 \km\ out to the CI clump. A second velocity gradient is seen in the PV diagram,  Figure \ref{fig_4} (top right),  running  NE (6936 \km) to SW (7085 \km) along an axis  approximately aligned with  knots D and knot E   ($PA$  $\sim$ 244\degree). Knot identification letters follow  the \halpha\ knots  from \cite{gava95} as marked in Figure  \ref{fig2} (bottom left).  \textcolor{black}{This second PV diagram shows two velocity components, one at $\sim$ 7000 \km\ and the other at $\sim$ 7050 \km.  }}
 
 \textcolor{black}{The  HERA \hto\ spectra for the CO disk,  CI  and a  combined CO disk + CI  spectrum are shown in Figure \ref{fig_3}. The peak signal to noise (S/N) in the spectra are \textcolor{black}{6.8, 3.9, and 7.3}, respectively. The \hto\ velocity  of the CO disk + CI  ($V$ = \textcolor{black}{7030 $\pm$12\km)} agrees within the errors with the \hi\ velocity of 7019 $\pm$21 \km\  \textcolor{black}{(Paper I)} as does the HERA \hto\ W$_{20}$ \textcolor{black}{(240$\pm$25 \km)} and \hi\ W$_{20}$ (216 $\pm$41 \km), where W$_{20}$ is the full width of the line at 20\% of the peak \textcolor{black}{emission in the spectrum}. } 
% The \hto\ velocity gradient observed along the this axis aligns  with  the \textcolor{black}{infrared} disk \textcolor{black}{(\textcolor{black}{defined} in section \ref{discuss})} major axis (PA = 286\degree), which shows V$_{CO}$ increasing from the SE ($\sim$  6914\,\km)  to the NW ($\sim$ 7085\,\km). A second velocity gradient is seen in the PV diagram -- Figure \ref{fig_4} (top left)  running   NE (6936 \km) to SW (7085 \km) along an axis  approximately aligned with  knots E and knot D   (PA  $\sim$ 250\degree),  Figure \ref{fig_3} (top right)}. \textcolor{black}{Knot identification letters follow  the \halpha\ knots  from \cite{gava95} marked in Figure  \ref{fig2} (bottom left).}
 
%\textcolor{black}{In addition to perturbing the  galaxy's \hi, the interaction which \cg079 is undergoing,  has also significantly perturbed both the morphology and kinematics of its \hto.}

\begin{table*} 
\centering
\begin{minipage}{140mm}
\caption{\hto\  properties from HERA}
\label{hera}
\begin{tabular}{@{}llrrr}
\hline
Property&Unit& CO disk&\textcolor{black}{CI} &\textcolor{black}{Disk + CI}\\
\hline
L(CO)$_{2-1}$ [T$_\mathrm{mb}]$&\textcolor{black}{[K \km pc$^2$]}&\textcolor{black}{1.2 $\pm0.1 \times 10^8$}&\textcolor{black}{0.2 $\pm0.05 \times 10^8$}&\textcolor{black}{1.4 $\pm0.1 \times 10^8$}   \\
M(H$_2$)&[10$^9$ \msolar]&\textcolor{black}{0.7 $\pm0.1$}&\textcolor{black}{0.1 $\pm0.03$}&\textcolor{black}{0.8 $\pm0.1$}\\

%V$_\mathrm{hel}$ &[\km]&6989 $\pm$6&\textcolor{black}{7074 $\pm$14}&\textcolor{black}{7040 $\pm$12}   \\
%W$_{20}$ &[\km]&192 $\pm$12 &\textcolor{black}{150 $\pm$28} & \textcolor{black}{260 $\pm$23} \\
V$_\mathrm{hel}$ &[\km]&\textcolor{black}{7000 $\pm$13}&\textcolor{black}{7045 $\pm$31}& \textcolor{black}{7030 $\pm$12}  \\
W$_{20}$ &[\km]&\textcolor{black}{220 $\pm$26} &\textcolor{black}{210 $\pm$62} & \textcolor{black}{240 $\pm$25} \\
rms per channel [T$_\mathrm{mb}]$&[mK] &&&3.1   \\
\textcolor{black}{rms integrated map [T$_\mathrm{mb}$]} &[K \km]&&&\textcolor{black}{0.23}   \\
\hline
\end{tabular}
\end{minipage}
\end{table*}

 \section{DISCUSSION}
\label{discuss}
In the following we summarise the properties of \textcolor{black}{the gas and stellar components} of \cg079, based on available data, including  a  series of papers by Gavazzi and collaborators \citep[e.g.,][]{gava78,gava87,bosel94,gava95,gava01b, bosel14b}. We subsequently consider the  \textcolor{black}{scenarios} most likely to explain the accumulated observations in the  concluding remarks (section \ref{concl}.)
\subsection{The  old stellar component}
\label{oldstellar}

\textcolor{black}{ A combination of perturbed ISM  and an unperturbed old stellar population is the} key signature  of an ongoing ram pressure stripping interaction \citep[e.g.,][]{ken04}. Hubble Space Telescope imaging  of  ESO\,137-001 which is suffering strong ram pressure stripping  \citep{jachym14}  shows narrow trails of blue optical emission, aligned downstream of the optical disk,  as well as extra--planar dust  above the  inner part of the optical disk. So it is clear that \textcolor{black}{extra--planar} emission/absorption from dust mixed with the stripped gas  as well as emission from young stars can result from ram pressure stripping.  \textcolor{black}{In general infrared (IR) emission is a better tracer  of the old stellar population than optical emission.} But IR emission from a galaxy's  old stellar population can be  contaminated by dust re--emission following absorption of UV photons from recently formed high mass stars \textcolor{black}{and stellar emission from RGB stars with ages of 10 to 100 Myr}. \cite{bosel04} give the dust emission contamination at  \textcolor{black}{6.75 $\mu$m as} up to 80\% for Sc galaxies and 50\% for blue compact dwarfs. But as \cite{bosel04} note MIR dust re--emission is more intense in massive quiescent galaxies than in star forming metal poor ones.  \cg079 is both a dwarf and metal poor \citep[$Fe/H$ = -0.62][]{mouhcine11}. However, while contamination by dust re--emission  is significant at  mid infrared  (MIR) wavelengths (including the 6.75 $\mu$m and 15 $\mu$m) it is minimal in the  \textit{Spitzer}  3.6$\mu$m and 4.5$\mu$m \citep{fazio05,utomo2014,meidt2014}. Modelling by \cite{utomo2014} \textcolor{black}{indicates} the contamination \textcolor{black}{(including from polycyclic aromatic hydrocarbon)} remains  minimal (of the order of a few \%) at 3.6$\mu$m and 4.5$\mu$m even for galaxies with  \halpha\ equivalent widths $(EW)$  similar to \cg079 ($EW$(\halpha) =129 per GOLDMine). \textcolor{black}{  For a disk galaxy contamination  from young stellar emission (principally from red supergiants)  at $\sim$2.3$\mu$m   is estimated  at $\sim$3\%, but in star forming regions within the galaxy the contamination  can be as high as 33\%  \citep{rhoads1998}.  The  strength of the emission  from red supergiants  in the wavelength range of  2.3 $\mu$m and 4.5 $\mu$m  declines with increasing wavelength  \citep{baron2014}. This suggests that  red supergiant contamination at 4.5  $\mu$m  is likely to be  less than at 2.3  $\mu$m by a factor of a few. }  \textcolor{black}{Because of the non--availability of a  \textit{Spitzer} 3.6$\mu$m image we were unable to make the polycyclic aromatic hydrocarbon (PAH)  and young stellar emission contamination corrections developed by  \citep{meidt12,meidt2014}.  \textcolor{black}{ We conclude that in the absence of these corrections the 4.5$\mu$m band image provides the  best available,  tracer of the old stellar population. This band is  known to suffer some CO absorption \citep{meidt2014}. } }

\begin{figure*}
\begin{center}
\includegraphics[ angle=0,scale=0.5] {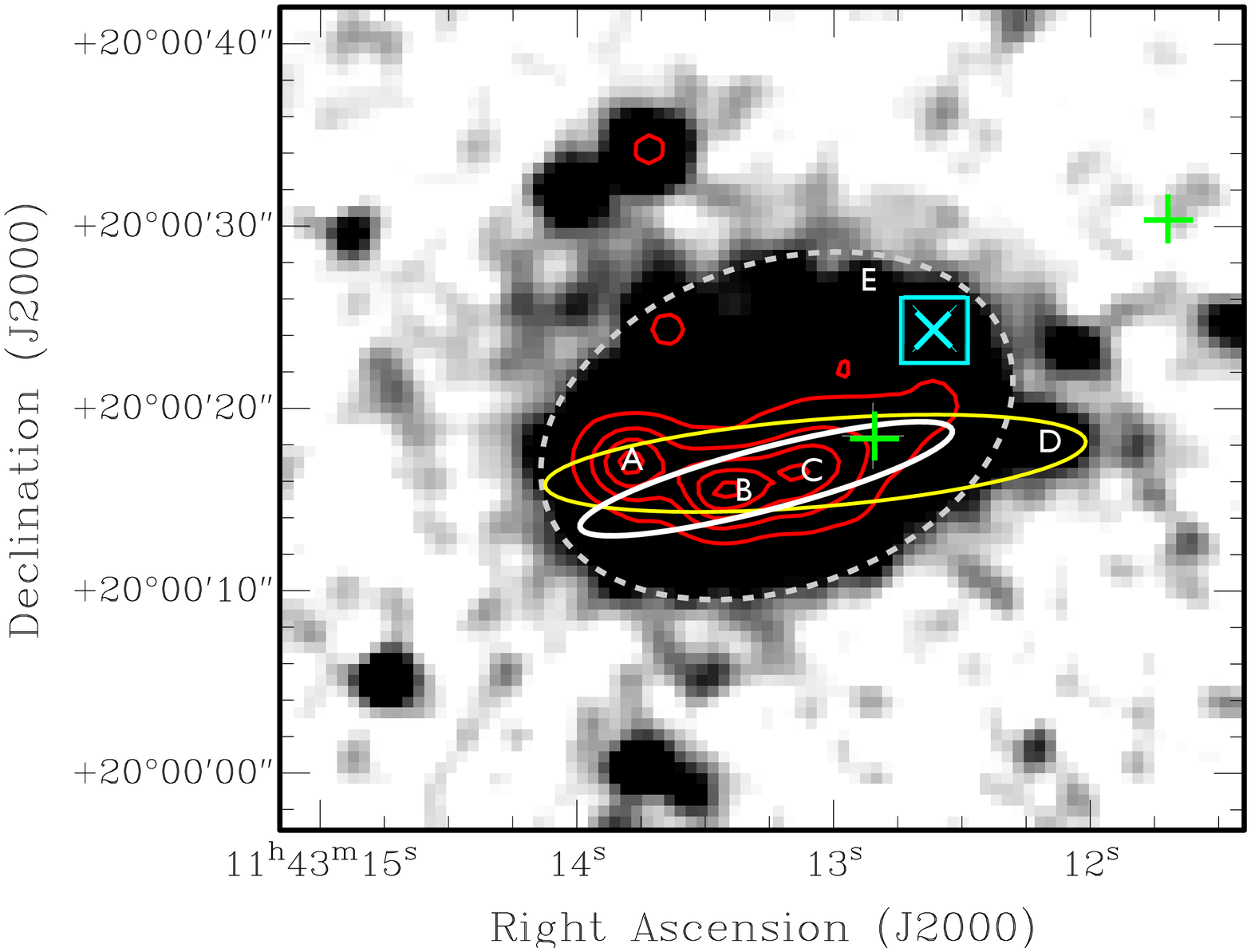}
\vspace{1cm}
\caption{ \textcolor{black}{ Stretched  4.5 $\mu$m  band image of \cg079. The  ellipses  indicate three  possible pre--interaction stellar disks:  the yellow ellipse represents a edge on, $\sim$ 90\degree\ inclination,  disk ($PA$ = 275\degree) encompassing knots  A, B, C and D;  the  white  dashed ellipse is a  lower inclination ($\sim$ 53\degree) inclination disk ($PA$ = 290\degree) tracing the outer edges of 4.5 $\mu$m -- band emission;  and the solid white  ellipse represents an edge on, $\sim$ 90\degree\ inclination,  disk ($PA$ = 286\degree) aligned with knots B and C. The cyan ’X’ in box marks the position of the \hi\ intensity maximum from the VLA C--array and the green plus symbols the \hto\ maximum and CI from the HERA map. The knot names are the same as used in Figure 2.   } }
\label{fig_6}
\end{center}
\end{figure*}

Figure~\ref{fig2} (bottom right) shows the \textit{Spitzer} 4.5 $\mu$m  image for \cg079, \textcolor{black}{ with contours from the same image with a 3--pixel wide  boxcar smoothing applied. The un--smoothed image had a pixel size = 1.2 arcsec and resolution $\sim$ 2.5 arsec). } Two principal intensity maxima are seen in the figure: an infrared counterpart to \halpha\ knot A, and a more elongated  counterpart approximately coinciding with \halpha\ knots B and C. The 4.5 $\mu$m intensity maximum is projected at approximately the same position as the optical centre and  knot B.  \textcolor{black}{The figure shows the intensity of the  4.5 $\mu$m emission falls steadily south of  \textcolor{black}{the} optical centre, but to the  N and NW of it there is an extensive region of diffuse emission. An uncatalogued  4.5 $\mu$m feature marked F on the figure is not present in  the 2MASS, optical (SDSS) or \halpha\ images, although it does have a  counterpart in our  $J$ -- band image obtained with the SPM\footnote{San Pedro M\'artir, Mexico (Observatorio Astron\'omico Nacional) } 2.1 m telescope.  Knot F's \textcolor{black}{colour suggests it is a higher redshift background galaxy. The optical counterparts of the two objects in the same figure  $\sim$ 10 arcsec N  of knot F are catalogued  as 23rd  magnitude stars in SDSS. }  Depending on the weighting given to the high and low  intensity  4.5 $\mu$m  emission there are  several possible \textcolor{black}{alternative  inclinations and morphologies for the old stellar disk}. Three  possible old stellar disk orientations are  over plotted as ellipses in Figure \ref{fig_6}. The yellow ellipse represents an edge--on, $\sim$ 90\degree\ inclination  disk ($PA$ = 275\degree) aligned with knots A, B, C and D,  the dashed  white  ellipse is a  lower inclination (60\degree) disk ($PA$ = 286\degree) aligned with the outer edges of the detected 4.5 $\mu$m,  and the solid white ellipse represents an edge on, $\sim$ 90\degree\ inclination  disk ($PA$ = 286\degree) aligned with knots B and C.  In case of the  high inclination disk options  the  diffuse 4.5 $\mu$m  emission to the north of the optical centre, and particularly the upward curve at the western disk edge, would be evidence of a tidal perturbation. On the other hand for the lower inclination  (60\degree) disk the northern diffuse emission would just be part of the disk. Because of these and \textcolor{black}{further} alternative interpretations of the 4.5 $\mu$m data\textcolor{black}{, as well as our inability  to make the PAH and young stellar emission contamination corrections,} we were unable to definitively determine \textcolor{black}{whether the morphology of the old stellar disk is perturbed or not.}}%Unfortunately the 4.5 $\mu$m morphology does not provide us with clear picture of impact of the interaction on the old stellar population, with observed morphology being the result of tidal perturbation of the pre--interaction irregular morphology.  }%An alternative interpretation of the infrared morphology is that the arc of  knots (A to E) traces a disk edge  in projection. %Under this scenario it is unclear how the old stellar disk, as traced by the infrared emission,  could remain stable with  $\sim$ 60\% of its mass located  along half of the perimeter of the disk.}

% \textcolor{black}{The broad diffuse  tail in the NW and knot A's  discrete nature  can also be seen  in  Figure \ref{fig_3} (bottom panel), which shows the  normalised \textit{Spitzer}  4.5 $\mu$m emission integrated along the major axis of the infared disk. } \textcolor{black}{Figure \ref{fig_5}  -- left panel, confirms that the \textit{Spitzer} 4.5 $\mu$m emission and \halpha\ are not well correlated in the regions of diffuse   4.5 $\mu$m emission, indicating that in these regions, at least, the 4.5 $\mu$m emission is likely to be predominantly tracing the old stellar population. } Seen in a field galaxy, the \textcolor{black}{4.5 $\mu$m} morphology \textcolor{black}{of the diffuse features to the N and NW,} would likely  be interpreted as indicative of a minor merger. 
\subsection{HI and CO gas components}
\label{neutralgas}
\cg079 is particularly gas rich, with M(\hdos\ + \hi\ + He) $ =  \textcolor{black}{2.9} \times10^9$ \msolar\ ($M_\mathrm{HI} =1.3 \times 10^9$\,\msolar\ -- Paper I) and  taking $M_* = 1.2 \times 10^9$ \msolar\  gives  a gas fraction\footnote{Ratio of gas mass to the sum of gas and stellar masses.}, $f_\mathrm{gas}$ = \textcolor{black}{0.7}.  Its  $M($\hi$)/M_* $=1.25 is consistent with that found for dwarf \citep{zhang2012} and smaller late--type galaxies \textcolor{black}{ \citep{cort11}} like the LMC and M33. Based on the HERA observations the \hdos\ \textcolor{black}{excess}    \textcolor{black}{(including CI) is -0.50, in good} agreement with the  \hdos\ excess from Paper II (-0.64).  \textcolor{black}{ As discussed in Paper II, there are significant uncertainties both in the determination of the \hdos\ mass of individual galaxies and the calibration of  \hdos\ deficiencies from samples of unresolved CO observations, but  based on the available data we can, at a minimum, conclude that \cg079 is not deficient in molecular gas.  \cite{gava01b} estimated the mass of \halpha\ in the ionised tail as 9.6 $\times$ 10$^8$ \msolar, which could  account for almost all of the \hi\ deficiency  (0.25) reported for \cg079 in Paper I, provided the \halpha\ emission is from gas that was originally \hi\ stripped from the galaxy.} The \textcolor{black}{NW offset  of the  \hto\ distribution from the optical centre, presented in section \ref{results}, is further} illustrated in Figure~\ref{fig_4} (lower panel), which shows the  normalised  \hto\ emission integrated along the axis of knots B and C \textcolor{black}{in comparison with 4.5 $\mu$m emission along the same axis.  The displacement of the \hto\ from the highest density old stellar component of the galaxy is clearly seen in the figure. } 

\textcolor{black}{Figure~\ref{fig2} (top right) shows that the \hi\ intensity maximum from  \textcolor{black}{the} VLA C--array is offset by 7\,arcsec (3\,kpc)  north--west of the \hto\ maximum and  \textcolor{black}{has  a  column density derived from the}  VLA data (C+D--array) of $6.6 \times 10^{20}$\,atom\,cm$^{- 2}$ \citep[][]{hota07}. This offset is consistent with our deeper VLA  -- D array observation in Paper I.   The Hota VLA data (C+D--array) map (their figure 14) shows the  \hi\ velocities increasing from 6923 \km\ in the NE to 7117 \km\ in the SW along the knot  E to D axis.  A similar  \hto\  velocity gradient is observed along the same axis  and the \hto\ PV \textcolor{black}{diagram} for a cut along this axis indicates there are two velocity components (Section \ref{results}).  Also detailed in section \ref{results} is the  \hto\ velocity gradient between the optical centre and the \hto\ maximum with a decrease in the gradient  NW of the \hto\ maximum.}

Both the  offsets of the \hi\ and \hto\ maxima  from the optical/4.5 $\mu$m centre  and the  segregation of the high column density \hi\  and \hto\ maxima  contrast with NGC\,4522 and NGC\,4330   in the Virgo cluster \citep{voll08,voll12}, where the high column density \hi\ and CO remain projected at the  optical centre  during ram pressure stripping. \textcolor{black}{Although NGC\,4522 and NGC\,4330 are both significantly more massive than \cg079. }%The \hi\ (C--array) and \hto\ maxima  offsets, relative to the optical centre  (PA = 304\degree, PA = 284\degree respectively), are imperfectly aligned  with the  the long  \halpha\ tail (PA =295\degree\ textcolor{black}{see Figure \ref{overv}}).
\subsection{Recent star formation} 
\begin{figure}
\begin{center}
\includegraphics[ angle=0,scale=0.5] {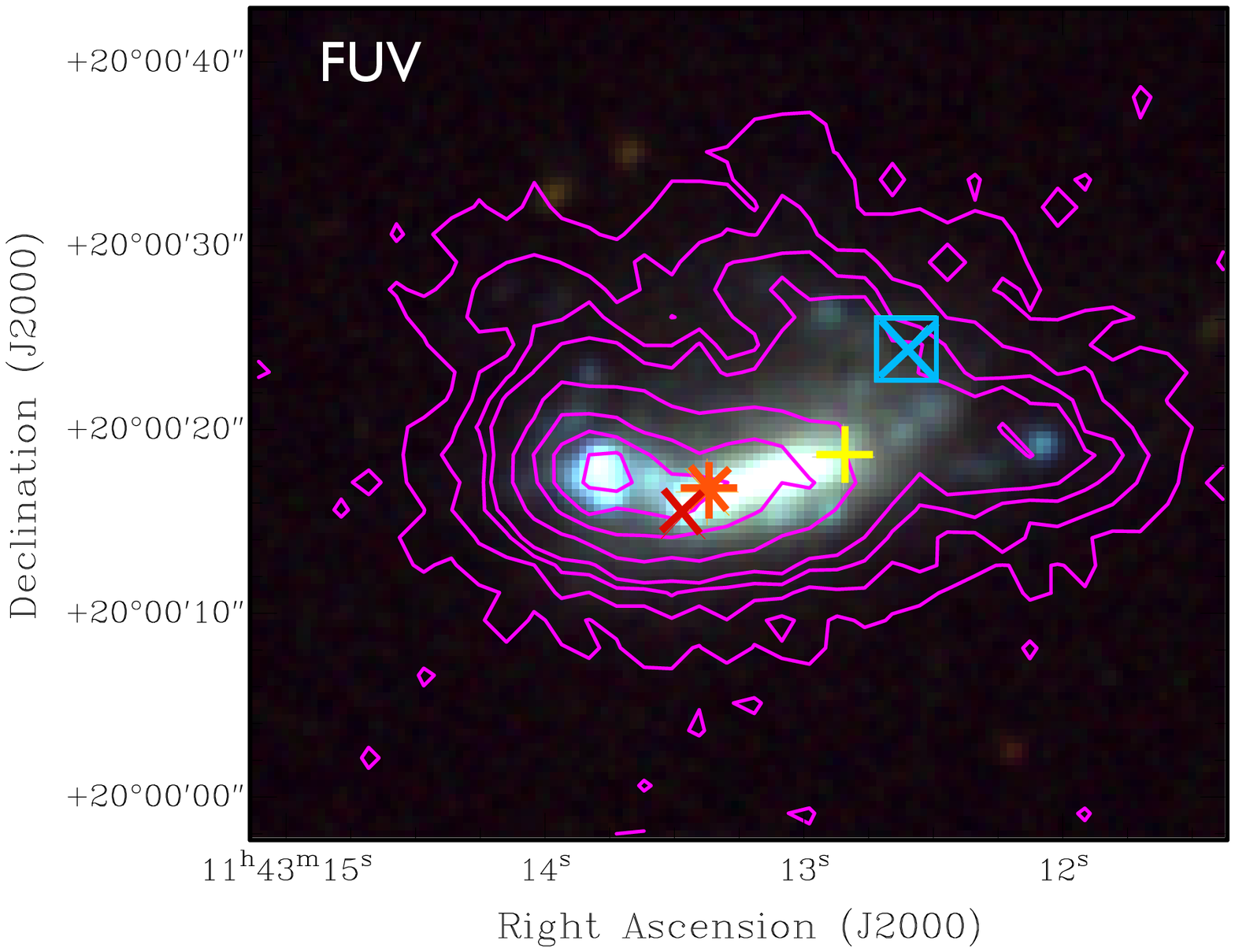}
\vspace{1cm}
\caption{ \textcolor{black}{ \cg0979: Contours from a FUV (\textit{GALEX})  image,  overlaid on  a composite  SDSS  false colour $g, r, i$ -- band image. The  contours are at 4	,15,	30, 45,	60,	100,	200,	300 and 	400
  $\sigma$ with the 4 $\sigma$ equivalent to  1.74 $\times 10^{-18}$   erg sec$^{-1}$ cm$^{-2}$  \angs $^{-1}$. The resolution of the FUV image is $\sim$ 4 arcsec. } }
\label{fig_7}
\end{center}
\end{figure}

\begin{figure*}
\begin{center}
\includegraphics[ angle=0,scale=0.5] {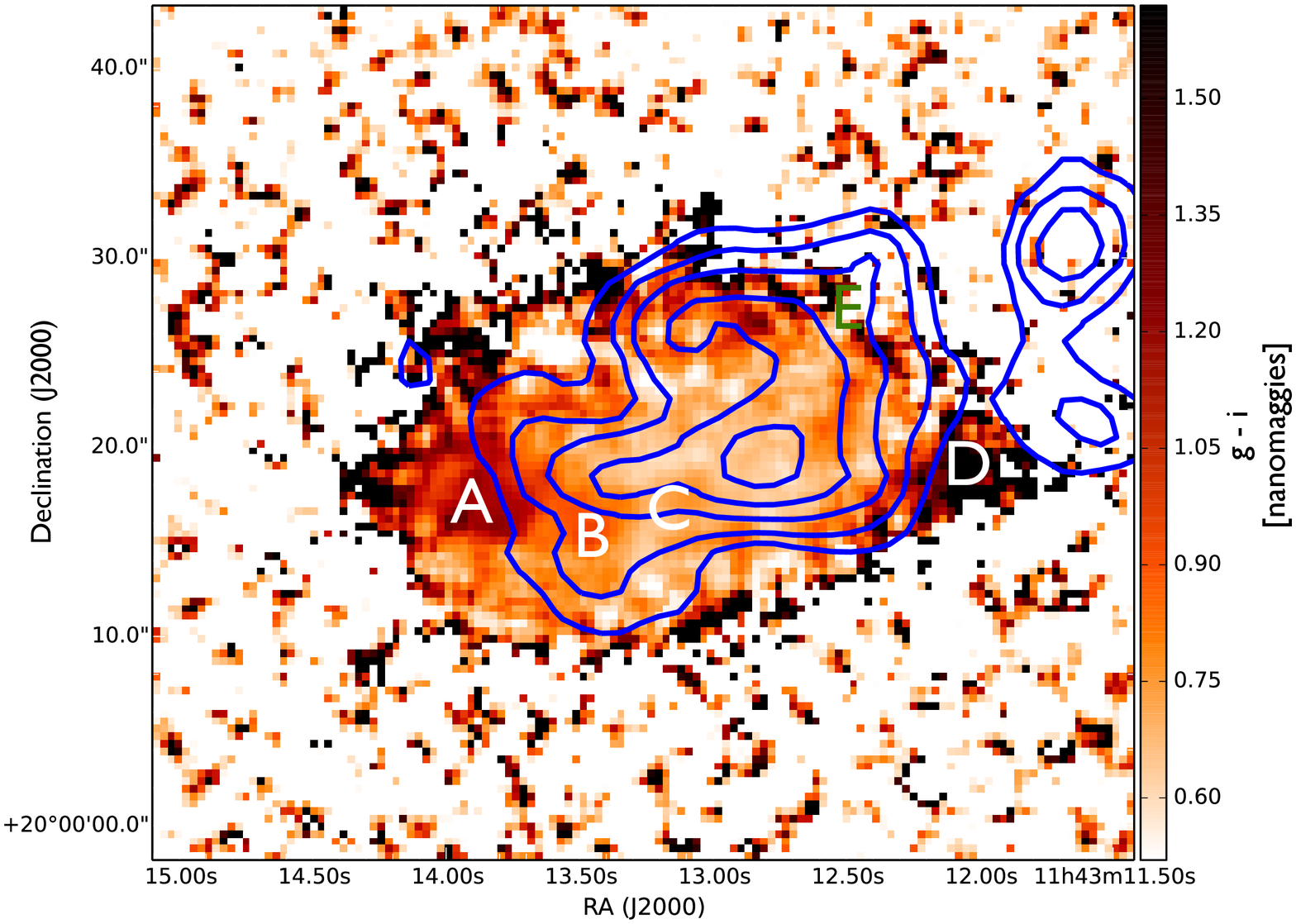}
\vspace{1cm}
\caption{\textcolor{black}{ \textbf{\cg079}   SSDS $g-i$  band image with  the same \hto\ contours \textcolor{black}{(blue) and \halpha\ knot identifications}  as in Figure 2. }   }
\label{fig_7_heat}
\end{center}
\end{figure*}

\textcolor{black}{ \halpha\ (Figure~\ref{fig2}, bottom left) and FUV \textit{GALEX} (Figure~\ref{fig_7}) emission trace star formation on time scales of 10$^7$\,yr and 10$^8$\,yr respectively \citep{bosel09} and are both concentrated east of the \hto\ maximum. About 30\% of the \cg079 FUV emission is  detected  within  a radius of \textcolor{black}{ $\sim$ }3 arcsec (1.25 kpc) of knot A. The \halpha\ emission  is concentrated in knots A, B and C. These  three knots are also the source of the strongest  1.5 GHz  radio continuum emission \citep[Figure 1b in][]{gava95}.  Figure \ref{fig_7_heat}  shows a $g - i$ band image produced from \textcolor{black}{SDSS} $g$ and $i$ band  background subtracted images overlaid with HERA \hto\ contours. From the  figure we see an extensive area between knots B and E with  $g - i$  $\sim$ 0.7. The figure shows this area correlates well with the high column density \hto\ emission. This strongly suggests the $g - i$ value in this region  is the result of dust extinction and the bluer $g - i$ colours for knots A and D are partially the result of lower extinction compared to knots B and C. Even allowing for this, knot D appears to be dominated by a younger  population of stars than  the four other knots.    An interesting feature in the \halpha\ image is the ridge of \halpha\  emission along the knot E to D axis ($PA$ = 244\degree) which  has less well defined optical (SDSS), 4.5 $\mu$m, \hi\  and \hto\ counterparts.  }

\subsection{High impact low velocity tidal interactions with  nearby neighbours } 
\label{neighbours}
\textcolor{black}{Here we consider whether there is a neighbouring galaxy or galaxies which could have tidally disrupted \cg079 within the last few $\times$ 10$^8$ yr assuming a typical group velocity of $\sim$ 250 \km. To try to identify any such galaxy we used } 
 a modified version of the tidal force ($Q$) and number density ($\eta = 2.5$) parameters from \cite{verley07}, applied to data from SDSS DR8. A \textcolor{black}{$Q < 0$}  indicates the gravitational forces within the galaxy exceed the gravitational force exerted by the neighbours. We find \textcolor{black}{$Q  = -1.10$} for  the 27 neighbours in the western sub--cloud of the NW sub--cluster \citep{cort04} and  $Q = -1.25$ for the nearest neighbour, A\,1367\,GP82\,1236 (see Figure~\ref{overv} at $RA$ = $11^{\mathrm h} 43^{\mathrm m}  10.961^{\mathrm s}$, and $DEC$ = $+20\degr 01^\prime 47.01^{\prime\prime}$).   The  tidal perturbation parameter\footnote{$p_{gg}= \frac{(M_{comp} / M_{gal}) }{(d/r_{gal})^3} $ where $M_{gal}$ and $M_{comp}$ are the masses of the galaxy  and companion respectively, $d$ is the separation and $r$ is the galaxy disk radius \citep{byrd90}, where values of $p_{gg} > 0.1$ likely lead to tidally induced star formation.} $p_{gg}$ = 0.004 \textcolor{black}{for \cg079 with A\,1367\,GP82\,1236} is more than an order of magnitude below that expected to produce detectable molecular gas perturbations in \cg079. We conclude from both the \textcolor{black}{$Q $ and $p_{gg}$} parameters and evidence that the perturbation  occurred within  $\sim10^8$\,yr \textcolor{black}{(section \ref{rps}), } that  the substantial  displacement of the  high column density molecular gas in \cg079 is unlikely to be due to a \textcolor{black}{ low velocity high impact} tidal interaction with A\,1367\,GP82\,1236.    

\subsection{\textcolor{black}{High velocity interactions}}
 \label{rps}
 \textcolor{black}{In this section we consider whether \cg079 may have undergone a recent high velocity ($\sim$ 1000 \km) hydrodynamic and/or tidal  interaction. }
If the  perturbation that caused \hi\ and CO maxima offsets also drove  the recent evolution of the stellar population we can infer the elapsed time since the interaction from the galaxy's integrated properties. GALEV evolutionary synthesis models \citep{kotu09}  with realistic parameters produce a set of solutions for the observed $M_* = 1.2 \times 10^9$ \msolar\ and $SFR$ =1.06 \msolar\,yr$^{-1}$ (Paper II) for star bursts  with burst strengths of $\sim$ 0.20 -- 0.25. The model constraint on these solutions from the optical and NIR colours is rather poor, using either the GALEV or FUV based Milky Way + internal  extinction, but favour solutions with  a time since the star burst  of  several $\times 10^7$ yr to a few $\times 10^8$ yr. The expected relaxation time scales for  perturbed  \hto\  and \hi\ \citep{holwerda11} together  with  GALEV modelling  indicate the perturbation of the galaxy  probably occurred  $\sim$ 10$^8$\,yr ago. 

\textcolor{black}{The  deep \halpha \ image of \cg079 and \cg073 in \cite[][their figure 21]{bosel14b} shows each galaxy to have  a spectacular  \halpha\  tail extending $\sim$ 100 kpc \textcolor{black}{and 70 kpc} from their respective optical disks and meeting in projection at $\sim$ 11:42:57.7, +20:02:29.42 (see Figure \ref{overv}).} At the \textcolor{black}{vertex of the}  \halpha\ tails  the complex \halpha\ morphology  strongly suggests a high velocity tidal encounter occurred between the two galaxies earlier in their infall to the cluster as proposed in \cite{gava01b}. \textcolor{black}{There are indications in \cite{bosel94} (at low significance) and \cite{sivan14} that the molecular gas in \cg073 is also offset in the direction of its tail. } Assuming the most recent star burst in \cg079 was driven by the interaction at the tail intersection between $\sim$ 7 $\times10 ^7$ yr to 2 $\times10^8$ yr ago (as indicated by the GALEV models),  the  \cg079 projected \halpha\ tail length  ($\sim$ 4.3\,arcmin -- 106\,kpc), implies a velocity in the plane of the sky in the range of 500 \km\ to 1500 \km\ in  agreement with the 1200 \km\ velocity  and 10$^{7.8}$ yr derived  by \cite{gava01b} for the survival time of the ionised tail.

Simple modelling  of ram pressure at \cg079's distance from the NW sub--cluster core (Paper~I) implies  $P_\mathrm{ram} \approx 10^{-11}$\,dyn\,cm$^{-2}$ for  \textcolor{black}{a} $V_\mathrm{rel}$  of  1000\,\km.  \cg079 \textcolor{black}{remains gas rich   ($f_\mathrm{gas}$ = \textcolor{black}{0.7})  with   no overall gas deficiency (although marginally \hi\ deficient)}  despite evidence from the \textcolor{black}{\halpha\ } tails that ram pressure stripping has operated for least $\sim$ 1 $\times$ 10$^8$  yr. This, the indication from section \ref{neutralgas} that most of the stripped \hi\ is now observed in the \halpha\ tails and the estimate of  P$_\mathrm{ram} \approx 10^{-11}$\,dyn\,cm$^{-2}$ all indicate  \cg079 has suffered only moderate ram pressure stripping. Further support for moderate ram pressure comes from the absence of  an \hi\ counterpart to the \halpha\ tail. Based on our  VLA -- D observations from Paper I we estimate  there could at most  be 3 $ \times 10^8$ \textcolor{black}{ \msolar\ } of \hi\  undetected in the tail, i.e. the \halpha\ + potential undetected \hi\ in the tail account for the \hi\ deficiency.

Ram pressure stripping models  have  usually been applied to  more massive and earlier Hubble--type spirals with classical bulges and lower gas fractions \citep[e.g., spirals with M$_* = 3.8 \times 10^{10}$ \msolar\   and $f_\mathrm{gas}$ = 0.08;][]{roed05}. At P$_\mathrm{ram} =  10^{-11}$\,dyn\,cm$^{-2}$ the presence/absence of a bulge does not impact the stripping efficiency \citep{sten12}.  But modelling of the ram pressure stripping of gas rich dwarves (M$_* = 0.6 \times 10^9$ \msolar) by  \cite{marco03}  have complete gas stripping time scales of  between $2 - 4 \times 10^8$ yr for ram pressures at least an order of magnitude below that predicted for \cg079 \textcolor{black}{ at V$_\mathrm{rel}$  = 1000\,\km.} 

\textcolor{black}{Confirmation of the ability of ram pressure stripping to remove almost all of the gas from a dwarf on timescales of a few $\times$ 10$^8$ yr comes from the study of IC 3418 \citep{ken14}. IC 3418 is a highly \hi\ deficient Irr dwarf  in Virgo projected 225 kpc from M87, with $\sim$ 1/3rd of the stellar mass of \cg079. The galaxy is thought to have undergone rapid ram pressure stripping within the last few $\times$ 10$^8$ yr which truncated its star formation. CO was marginally detected in the disk, which together evidence of  super giant starformation $\leq$ 10$^8$ yr ago in the stellar disk,   supports the proposition that the highest density gas remains within a dwarf's disk until ram pressure stripping is complete. \hi\ in IC\,3418 was only detected within the starforming tail.}

\textcolor{black}{ Given \cg079 has a stellar mass, a gas fraction and other properties which are similar to dwarves as well as the evidence ram pressure stripping has operated for at least 10$^8$ yr, we conclude that had  \cg079 been subject to continuous ram pressure much stronger than predicted from Paper I (say $P_\mathrm{ram}$ = $\sim1 \times  10^{-10}$\,dyn\,cm$^{-2}$) or its lower mass compared to generalised ram pressure models produced significantly more efficient \hi\ stripping, a substantial overall gas deficiency \textcolor{black}{and \hi\ tail} would be observable by now. Since \textcolor{black}{neither this deficiency nor an \hi\ tail is} observed,  steady ram pressure \textcolor{black}{seems unlikely to be the}   principal cause of the offsets between the  \hi, \hto\  and stellar intensity maxima. However  a sudden increase in ram pressure from a rapid increase in ICM density, such as that proposed for NGC\,4522 \citep{ken04}, might have \textcolor{black}{led} to the displacement of  the \hi\ and \hto\  with respect to the stellar maxima without sufficient time for gas deficiencies \textcolor{black}{or an \hi\ tail } to develop. Such an increase in ram pressure might arise from shocks or ICM substructure which are more likely in a merging cluster, although the  X--ray (\textit{XMM})  data does not show evidence  of an ICM  density enhancement at or close to \cg079. For the ram pressure stripping archetype  NGC\,4522 there is no indication that  the highest density \hto\ was displaced from its currently observed location (in a disk surrounding  the optical centre)  during the proposed spike in ram pressure. Although there is kinematic evidence that some \hi\ has fallen back subsequently  \citep{ken04}. The argument that  a similar order of magnitude  spike in ram pressure  is  the cause the observed separation of the highest density  \hto, \hi\ and stellar components in \cg079 would therefore  rely on  this being the consequence of the shallower potential in \cg079 compared to NGC \,4522.   }

%The  \halpha,   \textit{\textit{Spitzer IRAC}} 4.5 $\mu$m (Figure~\ref{fig2}) and \textit{FUV}and  radio continuum \citep{gava95} images all show emission  extending along the axis from knot A  to knot D (PA $\sim$ 270\degree relative to knot A) and in some cases beyond knot D.  \halpha,  \textit{FUV} and \textit{ Spitzer} 8 $\mu$m emission all show a ridge of recent star formation  along the  knot D--knot E axis. }
\section{Concluding remarks}
\label{concl}

\cg079 is a large gas rich dwarf with a  modest  \hi\ deficiency and a significant \hdos\  excess which  displays complex morphology (and gas kinematics) at each of the wavelengths considered above. Taken together  these clearly indicate  the galaxy has suffered a strong perturbation. This perturbation probably coincided with a burst of star formation which evolutionary synthesis models indicate occurred  $\sim$ 10$^8$ yr ago. 

\cg073 and \cg079  display impressive \textcolor{black}{$\sim$ 70 and 100 kpc \halpha\ tails respectively} in the deep \cite{bosel14b} \halpha\ image which meet each other in projection NW of \cg079.  Modelling in \cite{gava01b} suggests \textcolor{black}{ a  high-speed tidal interaction}, at the vertex of the  \halpha\ tails, subsequently enhanced the efficiency of ram pressure stripping of \hi\ from  both galaxies.  The  \halpha\ tails of both galaxies only become visible in the deep \cite{bosel14b} \halpha\ image from the tails' vertex onwards. Prior to the \textcolor{black}{high--speed tidal interaction}, ram pressure stripping was insufficient to produce detectable  \halpha\ tails, i.e., both tails are detectable  because the \textcolor{black}{ high--speed interaction}  `` loosened ''  gas in the disks, which following stripping and ionisation now present as the  \halpha\ tails. The time scale for the \textcolor{black}{high--speed interaction} assuming a V$_\mathrm{rel}$  of 1000 \km\ (which implies a current  P$_\mathrm{ram} =  10^{-11}$\,dyn\,cm$^{-2}$)  is 10$^8$ yr in good agreement with the  time since the last burst of star formation and the timescale from the \cite{gava01b} modeling.  Based on the analysis in section \ref{neighbours} we concluded that \hi, \hto\  and stellar intensity maxima offsets (maxima offsets) are not the result of a high impact low velocity interaction with a near neighbour. We also investigated whether the maxima offsets could be the result a recent minor  merger  but did not find convincing evidence for this. 
\textcolor{black}{A further possibility is the ram pressure stripping of gas has caused a shift in the cusp of the dark matter halo away from  centre of the dark matter (DM) halo via drag forces. This effect was modelled for a  medium mass (total mass = 10$^{10}$ \msolar) dwarf subject to \textcolor{black}{face--on} ram pressure stripping by   \citep{smith12}. Their model with \textcolor{black}{a similar $f_\mathrm{gas}$ (0.5)} and ram pressure to that calculated for \cg079  predicts a shift in the cusp and stellar disk centres from the centre of the DM halo of $<$ 0.5 kpc on time scales of 10$^8$ yr. But more critically in their  models the  gas within the truncation radius remains bound to the deepest potential of the DM, i.e., this mechanism is unlikely to explain the high density offsets. } The most likely remaining possibilities are   that the  offsets in neural and molecular ISM are the result of ram pressure and/or were produced by the \textcolor{black}{high--speed interaction} \textcolor{black}{with \cg073}. As we note in the introduction models and  observations of spirals more massive than \cg079 indicate the estimated ram pressure of P$_\mathrm{ram} =  10^{-11}$\,dyn\,cm$^{-2}$ alone would  not produce  such offsets. %, with the results from \cite{fosssati13}  suggesting the outside -- in gas stripping seen in larger spirals also occurs in dwarf galaxies. 
%However  it is hard to see how this could happen without simultaneously  producing a significant gas deficiency and detectable \hi\ tail, except under the spike in ram pressure scenario for which we have no evidence. 
Unlike ram pressure stripping, tidal interactions are known to be capable of  affecting  high density \hi\ and molecular gas located deep within the galactic gravitational potential well \citep{duc94,iono05} and the relaxation time scale for \hi\  is $>$ $1 \times 10^8$ yr \citep{holwerda11}.  High velocity "harassment" \textcolor{black}{type fly--by}  interactions in \textcolor{black}{higher mass} spirals than \cg079 are expected to have only a minimal impact on the  old stellar component \citep{duc08}, but  it is unclear whether this also the case for lower mass spirals  such as \cg079. \textcolor{black}{For the \cg079 / \cg073  we do not know whether the interaction between them was a fly--by or the disks penetrated each other. }%Moreover, the maxima offsets together with our estimate for the  timescale and ram pressure strength are consistent with the  The \halpha\  tails and both galaxies are projected with A\,1367's radio relic \cite{gava87}. Coincidently the \halpha\ tails vertex is near the NW edge of a radio relic, so an alternative explanation for the visibility  of the \halpha\  tails is that  relic provides high energy electrons to stimulate the \halpha\ emission. This could explain \halpha\ emission at the tail vertex  despite the expected \halpha\ cooling time of $\sim$ 10$^7$ yr being less than the time since the interaction ($\sim$ 10$^8$ yr). 

 In summary without clear evidence of the impact of the interaction on the old stellar disk (section  \ref{oldstellar}) and the uncertainly about whether the  shallower gravitational potential in \cg079, compared to the best studied cases,  could allow ram pressure or \textcolor{black}{a high--speed tidal interaction} to produce the observed maxima offsets, we are unable to determine whether ram pressure or \textcolor{black}{a high--speed tidal interaction}  was the principal cause of the observed maxima offsets. However ram pressure stripping is likely to be  playing a significant role in the perturbation of lower density  gas. Resolved ram pressure stripping  modellng  of  the ISM in spirals with masses similar to \cg079 would greatly assist in resolving this question.

\section*{Acknowledgments}
We are grateful to the anonymous referee for his/her helpful and insightful comments which have, significantly  improved the  paper. We would also like to thank Nicola Brassington, Elke Roediger and Martin Hardcastle for very useful discussions. LC acknowledges support under the Australian Research Council's Discovery Projects funding scheme (DP130100664). HBA acknowledges support for this project via CONACyT grant No.\ 50794. This research has made use of the NASA/IPAC Extragalactic Database (NED) which is operated by the Jet Propulsion Laboratory, California Institute of Technology, under contract with the National Aeronautics and Space Administration.

This research has made use of the Sloan Digital Sky Survey (SDSS). Funding for the SDSS and SDSS-II has been provided by the Alfred P. Sloan Foundation, the Participating Institutions, the National Science Foundation, the U.S. Department of Energy, the National Aeronautics and Space Administration, the Japanese Monbukagakusho, the Max Planck Society, and the Higher Education Funding Council for England. The SDSS Web Site is http://www.sdss.org/.

\bibliographystyle{mn2e}
\bibliography{cluster}

\begin{thebibliography}{}

\bibitem[\protect\citeauthoryear{{Baron}, {Monnier}, {Kiss}, {Neilson}, {Zhao}
  \& {Anderson}}{{Baron} et~al.}{2014}]{baron2014}
{Baron} F.,  {Monnier} J.~D.,  {Kiss} L.~L.,  {Neilson} H.~R.,  {Zhao} M.,
  {Anderson} M.,  2014, ApJ, 785, 46

\bibitem[\protect\citeauthoryear{{Boselli}, {Boissier}, {Cortese}, {Buat},
  {Hughes} \& {Gavazzi}}{{Boselli} et~al.}{2009}]{bosel09}
{Boselli} A.,  {Boissier} S.,  {Cortese} L.,  {Buat} V.,  {Hughes} T.~M.,
  {Gavazzi} G.,  2009, ApJ, 706, 1527

\bibitem[\protect\citeauthoryear{{Boselli}, {Cortese}, {Boquien}, {Boissier},
  {Catinella}, {Gavazzi}, {Lagos} \& {Saintonge}}{{Boselli}
  et~al.}{2014}]{bosel14}
{Boselli} A.,  {Cortese} L.,  {Boquien} M.,  {Boissier} S.,  {Catinella} B.,
  {Gavazzi} G.,  {Lagos} C.,    {Saintonge} A.,  2014, A\&A, 564, A67

\bibitem[\protect\citeauthoryear{{Boselli} \& {Gavazzi}}{{Boselli} \&
  {Gavazzi}}{2006}]{bosel06a}
{Boselli} A.,  {Gavazzi} G.,  2006, PASP, 118, 517

\bibitem[\protect\citeauthoryear{{Boselli} \& {Gavazzi}}{{Boselli} \&
  {Gavazzi}}{2014}]{bosel14b}
{Boselli} A.,  {Gavazzi} G.,  2014, 22, 74

\bibitem[\protect\citeauthoryear{{Boselli}, {Gavazzi}, {Combes} \&
  {Lequeux}}{{Boselli} et~al.}{1994}]{bosel94}
{Boselli} A.,  {Gavazzi} G.,  {Combes} F.,    {Lequeux} J.,  1994, A\&A, 285,
  69

\bibitem[\protect\citeauthoryear{{Boselli}, {Lequeux} \& {Gavazzi}}{{Boselli}
  et~al.}{2004}]{bosel04}
{Boselli} A.,  {Lequeux} J.,    {Gavazzi} G.,  2004, A\&A, 428, 409

\bibitem[\protect\citeauthoryear{{Byrd} \& {Valtonen}}{{Byrd} \&
  {Valtonen}}{1990}]{byrd90}
{Byrd} G.,  {Valtonen} M.,  1990, ApJ, 350, 89

\bibitem[\protect\citeauthoryear{{Chung}, {van Gorkom}, {Kenney}, {Crowl} \&
  {Vollmer}}{{Chung} et~al.}{2009}]{chung09}
{Chung} A.,  {van Gorkom} J.~H.,  {Kenney} J.~D.~P.,  {Crowl} H.,    {Vollmer}
  B.,  2009, AJ, 138, 1741

\bibitem[\protect\citeauthoryear{{Cortese}, {Boissier}, {Boselli}, {Bendo},
  {Buat}, {Davies}, {Eales}, {Heinis}, {Isaak} \& {Madden}}{{Cortese}
  et~al.}{2012}]{cort12}
{Cortese} L.,  {Boissier} S.,  {Boselli} A.,  {Bendo} G.~J.,  {Buat} V.,
  {Davies} J.~I.,  {Eales} S.,  {Heinis} S.,  {Isaak} K.~G.,    {Madden} S.~C.,
   2012, A\&A, 544, A101

\bibitem[\protect\citeauthoryear{{Cortese}, {Catinella}, {Boissier}, {Boselli}
  \& {Heinis}}{{Cortese} et~al.}{2011}]{cort11}
{Cortese} L.,  {Catinella} B.,  {Boissier} S.,  {Boselli} A.,    {Heinis} S.,
  2011, MNRAS, 415, 1797

\bibitem[\protect\citeauthoryear{{Cortese}, {Davies}, {Pohlen}, {Baes},
  {Bendo}, {Bianchi} \& {Boselli}}{{Cortese} et~al.}{2010}]{cort10}
{Cortese} L.,  {Davies} J.~I.,  {Pohlen} M.,  {Baes} M.,  {Bendo} G.~J.,
  {Bianchi} S.,    {Boselli} A.,  2010, A\&A, 518, L49

\bibitem[\protect\citeauthoryear{{Cortese}, {Gavazzi}, {Boselli},
  {Iglesias-Paramo} \& {Carrasco}}{{Cortese} et~al.}{2004}]{cort04}
{Cortese} L.,  {Gavazzi} G.,  {Boselli} A.,  {Iglesias-Paramo} J.,
  {Carrasco} L.,  2004, A\&A, 425, 429

\bibitem[\protect\citeauthoryear{{Donnelly}, {Markevitch}, {Forman}, {Jones},
  {David}, {Churazov} \& {Gilfanov}}{{Donnelly} et~al.}{1998}]{don98}
{Donnelly} R.~H.,  {Markevitch} M.,  {Forman} W.,  {Jones} C.,  {David} L.~P.,
  {Churazov} E.,    {Gilfanov} M.,  1998, ApJ, 500, 138

\bibitem[\protect\citeauthoryear{{Duc} \& {Bournaud}}{{Duc} \&
  {Bournaud}}{2008}]{duc08}
{Duc} P.-A.,  {Bournaud} F.,  2008, ApJ, 673, 787

\bibitem[\protect\citeauthoryear{{Duc} \& {Mirabel}}{{Duc} \&
  {Mirabel}}{1994}]{duc94}
{Duc} P.-A.,  {Mirabel} I.~F.,  1994, A\&A, 289, 83

\bibitem[\protect\citeauthoryear{{Fazio}}{{Fazio}}{2005}]{fazio05}
{Fazio} G.~G.,  2005, in {Shapiro} M.~M.,  {Stanev} T.,   {Wefel} J.~P.,  eds,
  Neutrinos and Explosive Events in the Universe {Recent Results from the
  Spitzer Space Telescope: A New View of the Infrared Universe}.
p.~47

\bibitem[\protect\citeauthoryear{{Fossati}, {Gavazzi}, {Savorgnan},
  {Fumagalli}, {Boselli}, {Guti{\'e}rrez}, {Hern{\'a}ndez Toledo}, {Giovanelli}
  \& {Haynes}}{{Fossati} et~al.}{2013}]{fosssati13}
{Fossati} M.,  {Gavazzi} G.,  {Savorgnan} G.,  {Fumagalli} M.,  {Boselli} A.,
  {Guti{\'e}rrez} L.,  {Hern{\'a}ndez Toledo} H.,  {Giovanelli} R.,    {Haynes}
  M.~P.,  2013, A\&A, 553, A91

\bibitem[\protect\citeauthoryear{{Fumagalli}, {Krumholz}, {Prochaska},
  {Gavazzi} \& {Boselli}}{{Fumagalli} et~al.}{2009}]{fuma09}
{Fumagalli} M.,  {Krumholz} M.~R.,  {Prochaska} J.~X.,  {Gavazzi} G.,
  {Boselli} A.,  2009, ApJ, 697, 1811

\bibitem[\protect\citeauthoryear{{Gavazzi}}{{Gavazzi}}{1978}]{gava78}
{Gavazzi} G.,  1978, A\&A, 69, 355

\bibitem[\protect\citeauthoryear{{Gavazzi}, {Boselli}, {Donati}, {Franzetti} \&
  {Scodeggio}}{{Gavazzi} et~al.}{2003}]{gava03b}
{Gavazzi} G.,  {Boselli} A.,  {Donati} A.,  {Franzetti} P.,    {Scodeggio} M.,
  2003, A\&A, 400, 451

\bibitem[\protect\citeauthoryear{{Gavazzi}, {Boselli}, {Mayer},
  {Iglesias-Paramo}, {V{\'{\i}}lchez} \& {Carrasco}}{{Gavazzi}
  et~al.}{2001}]{gava01b}
{Gavazzi} G.,  {Boselli} A.,  {Mayer} L.,  {Iglesias-Paramo} J.,
  {V{\'{\i}}lchez} J.~M.,    {Carrasco} L.,  2001, ApJL, 563, L23

\bibitem[\protect\citeauthoryear{{Gavazzi}, {Contursi}, {Carrasco}, {Boselli},
  {Kennicutt}, {Scodeggio} \& {Jaffe}}{{Gavazzi} et~al.}{1995}]{gava95}
{Gavazzi} G.,  {Contursi} A.,  {Carrasco} L.,  {Boselli} A.,  {Kennicutt} R.,
  {Scodeggio} M.,    {Jaffe} W.,  1995, A\&A, 304, 325

\bibitem[\protect\citeauthoryear{{Gavazzi} \& {Jaffe}}{{Gavazzi} \&
  {Jaffe}}{1987}]{gava87}
{Gavazzi} G.,  {Jaffe} W.,  1987, A\&A, 186, L1

\bibitem[\protect\citeauthoryear{{Holwerda}, {Pirzkal}, {Cox}, {de Blok},
  {Weniger}, {Bouchard}, {Blyth} \& {van der Heyden}}{{Holwerda}
  et~al.}{2011}]{holwerda11}
{Holwerda} B.~W.,  {Pirzkal} N.,  {Cox} T.~J.,  {de Blok} W.~J.~G.,  {Weniger}
  J.,  {Bouchard} A.,  {Blyth} S.-L.,    {van der Heyden} K.~J.,  2011, MNRAS,
  416, 2426

\bibitem[\protect\citeauthoryear{{Hota} \& {Saikia}}{{Hota} \&
  {Saikia}}{2007}]{hota07}
{Hota} A.,  {Saikia} D.~J.,  2007, Bulletin of the Astronomical Society of
  India, 35, 121

\bibitem[\protect\citeauthoryear{{Iono}, {Yun} \& {Ho}}{{Iono}
  et~al.}{2005}]{iono05}
{Iono} D.,  {Yun} M.~S.,    {Ho} P.~T.~P.,  2005, ApJS, 158, 1

\bibitem[\protect\citeauthoryear{{J{\'a}chym}, {Combes}, {Cortese}, {Sun} \&
  {Kenney}}{{J{\'a}chym} et~al.}{2014}]{jachym14}
{J{\'a}chym} P.,  {Combes} F.,  {Cortese} L.,  {Sun} M.,    {Kenney} J.~D.~P.,
  2014, ApJ, 792, 11

\bibitem[\protect\citeauthoryear{{Kapferer}, {Sluka}, {Schindler}, {Ferrari} \&
  {Ziegler}}{{Kapferer} et~al.}{2009}]{kapf09}
{Kapferer} W.,  {Sluka} C.,  {Schindler} S.,  {Ferrari} C.,    {Ziegler} B.,
  2009, A\&A, 499, 87

\bibitem[\protect\citeauthoryear{{Kenney}, {Geha}, {J{\'a}chym}, {Crowl},
  {Dague}, {Chung}, {van Gorkom} \& {Vollmer}}{{Kenney} et~al.}{2014}]{ken14}
{Kenney} J.~D.~P.,  {Geha} M.,  {J{\'a}chym} P.,  {Crowl} H.~H.,  {Dague} W.,
  {Chung} A.,  {van Gorkom} J.,    {Vollmer} B.,  2014, ApJ, 780, 119

\bibitem[\protect\citeauthoryear{{Kenney}, {van Gorkom} \& {Vollmer}}{{Kenney}
  et~al.}{2004}]{ken04}
{Kenney} J.~D.~P.,  {van Gorkom} J.~H.,    {Vollmer} B.,  2004, AJ, 127, 3361

\bibitem[\protect\citeauthoryear{{Koopmann} \& {Kenney}}{{Koopmann} \&
  {Kenney}}{2004}]{koop04}
{Koopmann} R.~A.,  {Kenney} J.~D.~P.,  2004, ApJ, 613, 851

\bibitem[\protect\citeauthoryear{{Kotulla}, {Fritze}, {Weilbacher} \&
  {Anders}}{{Kotulla} et~al.}{2009}]{kotu09}
{Kotulla} R.,  {Fritze} U.,  {Weilbacher} P.,    {Anders} P.,  2009, MNRAS,
  396, 462

\bibitem[\protect\citeauthoryear{{Leroy}, {Walter}, {Bigiel}, {Usero}, {Weiss},
  {Brinks}, {de Blok}, {Kennicutt}, {Schuster}, {Kramer}, {Wiesemeyer} \&
  {Roussel}}{{Leroy} et~al.}{2009}]{leroy09}
{Leroy} A.~K.,  {Walter} F.,  {Bigiel} F.,  {Usero} A.,  {Weiss} A.,  {Brinks}
  E.,  {de Blok} W.~J.~G.,  {Kennicutt} R.~C.,  {Schuster} K.-F.,  {Kramer} C.,
   {Wiesemeyer} H.~W.,    {Roussel} H.,  2009, AJ, 137, 4670

\bibitem[\protect\citeauthoryear{{Marcolini}, {Brighenti} \&
  {D'Ercole}}{{Marcolini} et~al.}{2003}]{marco03}
{Marcolini} A.,  {Brighenti} F.,    {D'Ercole} A.,  2003, MNRAS, 345, 1329

\bibitem[\protect\citeauthoryear{{Meidt}, {Schinnerer}, {Knapen}, {Bosma},
  {Athanassoula} \& {Sheth}}{{Meidt} et~al.}{2012}]{meidt12}
{Meidt} S.~E.,  {Schinnerer} E.,  {Knapen} J.~H.,  {Bosma} A.,  {Athanassoula}
  E.,    {Sheth} K.,  2012, ApJ, 744, 17

\bibitem[\protect\citeauthoryear{{Meidt}, {Schinnerer}, {van de Ven},
  {Zaritsky}, {Peletier}, {Knapen}, {Sheth} \& {Regan}}{{Meidt}
  et~al.}{2014}]{meidt2014}
{Meidt} S.~E.,  {Schinnerer} E.,  {van de Ven} G.,  {Zaritsky} D.,  {Peletier}
  R.,  {Knapen} J.~H.,  {Sheth} K.,    {Regan} M.,  2014, ApJ, 788, 144

\bibitem[\protect\citeauthoryear{{Mouhcine}, {Kriwattanawong} \&
  {James}}{{Mouhcine} et~al.}{2011}]{mouhcine11}
{Mouhcine} M.,  {Kriwattanawong} W.,    {James} P.~A.,  2011, MNRAS, 412, 1295

\bibitem[\protect\citeauthoryear{{Nishiyama}, {Nakai} \& {Kuno}}{{Nishiyama}
  et~al.}{2001}]{nishiyama01}
{Nishiyama} K.,  {Nakai} N.,    {Kuno} N.,  2001, PASJ, 53, 757

\bibitem[\protect\citeauthoryear{{Plionis}, {Tovmassian} \&
  {Andernach}}{{Plionis} et~al.}{2009}]{plionis09}
{Plionis} M.,  {Tovmassian} H.~M.,    {Andernach} H.,  2009, MNRAS, 395, 2

\bibitem[\protect\citeauthoryear{{Rhoads}}{{Rhoads}}{1998}]{rhoads1998}
{Rhoads} J.~E.,  1998, AJ, 115, 472

\bibitem[\protect\citeauthoryear{{Roediger} \& {Br{\"u}ggen}}{{Roediger} \&
  {Br{\"u}ggen}}{2007}]{roed07}
{Roediger} E.,  {Br{\"u}ggen} M.,  2007, MNRAS, 380, 1399

\bibitem[\protect\citeauthoryear{{Roediger} \& {Hensler}}{{Roediger} \&
  {Hensler}}{2005}]{roed05}
{Roediger} E.,  {Hensler} G.,  2005, A\&A, 433, 875

\bibitem[\protect\citeauthoryear{{Scott}, {Bravo-Alfaro}, {Brinks}, {Caretta},
  {Cortese}, {Boselli}, {Hardcastle}, {Croston} \& {Plauchu}}{{Scott}
  et~al.}{2010}]{scott10}
{Scott} T.~C.,  {Bravo-Alfaro} H.,  {Brinks} E.,  {Caretta} C.~A.,  {Cortese}
  L.,  {Boselli} A.,  {Hardcastle} M.~J.,  {Croston} J.~H.,    {Plauchu} I.,
  2010, MNRAS, 403, 1175

\bibitem[\protect\citeauthoryear{{Scott}, {Usero}, {Brinks}, {Boselli},
  {Cortese} \& {Bravo-Alfaro}}{{Scott} et~al.}{2013}]{scott13}
{Scott} T.~C.,  {Usero} A.,  {Brinks} E.,  {Boselli} A.,  {Cortese} L.,
  {Bravo-Alfaro} H.,  2013, MNRAS, 429, 221

\bibitem[\protect\citeauthoryear{{Sivanandam}, {Rieke} \& {Rieke}}{{Sivanandam}
  et~al.}{2014}]{sivan14}
{Sivanandam} S.,  {Rieke} M.~J.,    {Rieke} G.~H.,  2014, ApJ, 796, 89

\bibitem[\protect\citeauthoryear{{Smith}, {Fellhauer} \& {Assmann}}{{Smith}
  et~al.}{2012}]{smith12}
{Smith} R.,  {Fellhauer} M.,    {Assmann} P.,  2012, MNRAS, 420, 1990

\bibitem[\protect\citeauthoryear{{Spergel}, {Bean}, {Dor{\'e}}, {Nolta},
  {Bennett}, {Dunkley}, {Hinshaw}, {Jarosik}, {Komatsu}, {Page}, {Peiris},
  {Verde}, {Halpern}, {Hill}, {Kogut}, {Limon}, {Wollack} \&
  {Wright}}{{Spergel} et~al.}{2007}]{sperg07}
{Spergel} D.~N.,  {Bean} R.,  {Dor{\'e}} O.,  {Nolta} M.~R.,  {Bennett} C.~L.,
  {Dunkley} J.,  {Hinshaw} G.,  {Jarosik} N.,  {Komatsu} E.,  {Page} L.,
  {Peiris} H.~V.,  {Verde} L.,  {Halpern} M.,  {Hill} R.~S.,  {Kogut} A.,
  {Limon} M.,  {Wollack} E.,    {Wright} E.~L.,  2007, ApJS, 170, 377

\bibitem[\protect\citeauthoryear{{Steinhauser}, {Haider}, {Kapferer} \&
  {Schindler}}{{Steinhauser} et~al.}{2012}]{sten12}
{Steinhauser} D.,  {Haider} M.,  {Kapferer} W.,    {Schindler} S.,  2012, A\&A,
  544, A54

\bibitem[\protect\citeauthoryear{{Tonnesen} \& {Bryan}}{{Tonnesen} \&
  {Bryan}}{2009}]{tonn09}
{Tonnesen} S.,  {Bryan} G.~L.,  2009, ApJ, 694, 789

\bibitem[\protect\citeauthoryear{{Utomo}, {Kriek}, {Labb{\'e}}, {Conroy} \&
  {Fumagalli}}{{Utomo} et~al.}{2014}]{utomo2014}
{Utomo} D.,  {Kriek} M.,  {Labb{\'e}} I.,  {Conroy} C.,    {Fumagalli} M.,
  2014, ApJL, 783, L30

\bibitem[\protect\citeauthoryear{{van Gorkom}}{{van Gorkom}}{2004}]{vgork04}
{van Gorkom} J.~H.,  2004, in {Mulchaey} J.~S.,  {Dressler} A.,   {Oemler} A.,
  eds, Clusters of Galaxies: Probes of Cosmological Structure and Galaxy
  Evolution {Interaction of Galaxies with the Intracluster Medium}.
p.~305

\bibitem[\protect\citeauthoryear{{Verley}, {Leon}, {Verdes-Montenegro},
  {Combes}, {Sabater}, {Sulentic}, {Bergond}, {Espada}, {Garc{\'{\i}}a},
  {Lisenfeld} \& {Odewahn}}{{Verley} et~al.}{2007}]{verley07}
{Verley} S.,  {Leon} S.,  {Verdes-Montenegro} L.,  {Combes} F.,  {Sabater} J.,
  {Sulentic} J.,  {Bergond} G.,  {Espada} D.,  {Garc{\'{\i}}a} E.,  {Lisenfeld}
  U.,    {Odewahn} S.~C.,  2007, A\&A, 472, 121

\bibitem[\protect\citeauthoryear{{Vollmer}, {Braine}, {Balkowski}, {Cayatte} \&
  {Duschl}}{{Vollmer} et~al.}{2001}]{voll01b}
{Vollmer} B.,  {Braine} J.,  {Balkowski} C.,  {Cayatte} V.,    {Duschl} W.~J.,
  2001, A\&A, 374, 824

\bibitem[\protect\citeauthoryear{{Vollmer}, {Braine}, {Pappalardo} \&
  {Hily-Blant}}{{Vollmer} et~al.}{2008}]{voll08}
{Vollmer} B.,  {Braine} J.,  {Pappalardo} C.,    {Hily-Blant} P.,  2008, A\&A,
  491, 455

\bibitem[\protect\citeauthoryear{{Vollmer}, {Soida}, {Braine}, {Abramson},
  {Beck}, {Chung}, {Crowl}, {Kenney} \& {van Gorkom}}{{Vollmer}
  et~al.}{2012}]{voll12}
{Vollmer} B.,  {Soida} M.,  {Braine} J.,  {Abramson} A.,  {Beck} R.,  {Chung}
  A.,  {Crowl} H.~H.,  {Kenney} J.~D.~P.,    {van Gorkom} J.~H.,  2012, A\&A,
  537, A143

\bibitem[\protect\citeauthoryear{{Zhang}, {Hunter}, {Elmegreen}, {Gao} \&
  {Schruba}}{{Zhang} et~al.}{2012}]{zhang2012}
{Zhang} H.-X.,  {Hunter} D.~A.,  {Elmegreen} B.~G.,  {Gao} Y.,    {Schruba} A.,
   2012, AJ, 143, 47

\end{thebibliography}

\end{document}